\documentclass[a4paper,fleqn]{cas-dc}

\usepackage[authoryear]{natbib}
\usepackage[T1]{fontenc}
\usepackage[utf8]{inputenc}

\def\tsc#1{\csdef{#1}{\textsc{\lowercase{#1}}\xspace}}
\tsc{WGM}
\tsc{QE}

\usepackage{subcaption}
\usepackage{microtype}

\usepackage{amssymb}
\usepackage{amsmath}

\usepackage{multirow}
\usepackage{booktabs}
\usepackage[dvipsnames]{xcolor}
\usepackage[export]{adjustbox}
\usepackage{enumitem}
\usepackage{csquotes}
\usepackage{hyperref}

\begin{document}

\let\WriteBookmarks\relax
\def\floatpagepagefraction{1}
\def\textpagefraction{.001}

\shorttitle{Learned Nonlocal Feature Matching and Filtering for RAW Image Denoising}
\title[mode=title]{Learned Nonlocal Feature Matching and Filtering for RAW Image Denoising}

\shortauthors{Sánchez-Beeckman and Buades}
\author[uib,iac3]{Marco Sánchez-Beeckman}[orcid=0000-0002-5949-0775]
\ead{marco.sanchez@uib.es}
\cormark[1]
\credit{Methodology, Software, Validation, Formal analysis, Investigation, Data Curation, Writing - Original Draft, Writing - Review \& Editing, Visualization}

\author[uib,iac3]{Antoni Buades}[orcid=0000-0001-9832-3358]
\ead{toni.buades@uib.es}
\credit{Conceptualization, Writing - Review \& Editing, Supervision}

\affiliation[uib]{organization={Dept. of Mathematics and Computer Science, Universitat de les Illes Balears},
            addressline={Cra.\ de Valldemossa km 7.5}, 
            city={Palma},
            postcode={07122}, 
            state={Illes Balears},
            country={Spain}}

\affiliation[iac3]{organization={Institute of Applied Computing and Community Code (IAC3), Universitat de les Illes Balears},
            addressline={C/ Blaise Pascal 7, Parc BIT}, 
            city={Palma},
            postcode={07121}, 
            state={Illes Balears},
            country={Spain}}

\cortext[1]{Corresponding author}

\begin{abstract}
Being one of the oldest and most basic problems in image processing, image denoising has seen a resurgence spurred by rapid advances in deep learning.
Yet, most modern denoising architectures make limited use of the technical knowledge acquired researching the classical denoisers that came before the mainstream use of neural networks, instead relying on depth and large parameter counts.
This poses a challenge not only for understanding the properties of such networks, but also for deploying them on real devices which may present resource constraints and diverse noise profiles.
Tackling both issues, we propose an architecture dedicated to RAW-to-RAW denoising that incorporates the interpretable structure of classical self-similarity-based denoisers into a fully learnable neural network. 
Our design centers on a novel nonlocal block that parallels the established pipeline of neighbor matching, collaborative filtering and aggregation popularized by nonlocal patch-based methods, operating on learned multiscale feature representations.
This built-in nonlocality efficiently expands the receptive field, sufficing a single block per scale with a moderate number of neighbors to obtain high-quality results.
Training the network on a curated dataset with clean real RAW data and modeled synthetic noise while conditioning it on a noise level map yields a sensor-agnostic denoiser that generalizes effectively to unseen devices.
Both quantitative and visual results on benchmarks and in-the-wild photographs position our method as a practical and interpretable solution for real-world RAW denoising, achieving results competitive with state-of-the-art convolutional and transformer-based denoisers while using significantly fewer parameters.
The code is available at \url{https://github.com/MIA-UIB/nonlocal-matchfilter}.
\end{abstract}



\maketitle

\section{Introduction}
\label{sec:intro}

The presence of noise in digital images is inevitable due to the very own workings of the image formation process.
Camera sensors, as well as more complex image acquisition systems, 
are susceptible to undesirable measurement errors that arise from multiple sources.
These include the variability caused by the random nature of photon counting, thermal fluctuations in the electronic circuits of the sensors, and conversion to digital signals~\citep{Wei2022Physics}, among other setting-specific perturbations. 
While modern advances in sensor technology have allowed for a reduction of noise magnitude at the physical level~\citep{Boukhayma2016NoiseCMOS}, it still remains a fundamental problem that must be dealt with after image capture.

Noise obscures fine details and textures, deteriorating signal information.
In a world where visual quality is paramount for phone camera users, its presence in day-to-day pictures is not acceptable, except for specific artistic purposes.
It also hinders interpretation in diagnostic-oriented images (e.g.\ medical images), and compromises the performance of downstream computer vision tasks~\citep{Pei2021DegradationEffects}.
For all these reasons, effective denoising methods that faithfully recover the underlying scene structure become essential for ensuring that images meet quality standards expected in modern imaging applications.

Over the years, a significant share of work has been notably dedicated to studying the removal of Additive White Gaussian Noise (AWGN)~\citep{Elad2023Survey}, whose simplicity makes it an ideal testbed for denoising methods.
In modern digital photography, however, denoising must address noise characteristics that differ substantially from the AWGN model.
At the sensor level, the combination of the two primary sources of noise (shot and readout noise) follows a compound Poisson-Gaussian distribution.
Its approximation by a heteroskedastic Gaussian with signal-dependent variance is often used as the simplest noise model for RAW images~\citep{Healey1994CCD,Foi2008NoiseModeling}, with more complex approaches proposed for low-light scenarios~\citep{Zhang2021Rethinking,Wei2022Physics,Zhang2023General,Cao2023Physics,Feng2024LearnabilityEnhancement,Lu2025DarkNoise}.
Critically, noise parameters in such models can vary significantly across devices and camera settings.
This motivates the use of noise-aware RAW denoisers that use prior noise level information to adapt to sensor characteristics.

Denoising algorithms can be broadly categorized into model-based and data-driven approaches.
Classical model-based methods rely on meticulously hand-crafted priors that capture natural image statistics.
Among these, patch-based methods exploiting nonlocal self-similarity have had the most influence in the field~\citep{Buades2005NLMeans}.
Perhaps the most widely known is the seminal BM3D algorithm~\citep{Dabov2007BM3D}, which spearheaded a family of denoisers with a common three-step pipeline: matching similar patches, transforming them into a suitable domain for filtering, and aggregating them back.
While these methods usually assume Gaussian noise, using its variance to adjust the strength of the filtering, they can be extended to the sensor-specific distributions of the RAW domain by using variance stabilization transforms~\citep{Makitalo2014NoiseMismatch} or local variance estimators~\citep{Sanchez2025Combining}.
Although they have been overshadowed by neural networks, classical methods remain relevant today, as their strong theoretical foundations and explicit structure provide insight into how different image features are processed.

In contrast, data-driven approaches based on deep learning, beginning with convolutional neural networks (CNNs) like DnCNN~\citep{Zhang2017DnCNN} and later with transformers~\citep{Liang2021SwinIR},
have shown excellent performance by learning to denoise directly from large datasets, up to the point of surpassing classical methods in quality.
By accepting a noise level map as input, networks like FFDNet~\citep{Zhang2018FFDNet} and DRUNet~\citep{Zhang2022DPIR} allow some degree of control over the filtering strength, an idea that has also been applied with success in the RAW domain~\citep{Li2024DualDn}.
However, even these noise-aware networks remain largely opaque, as their deep end-to-end learned representations obscure how they use image structures to remove noise.

This opacity is partly rooted in the architectural choices that uphold most modern denoisers, which use little domain knowledge cultivated before the ubiquity of deep learning.
Convolutional networks, whose layers operate locally by design, cannot exploit the nonlocal self-similarity that proved so effective in classical methods.
Transformers, via their attention mechanism, do perform data-dependent weighted averaging that evokes classical nonlocal adaptive filters~\citep{Milanfar2013Tour},
yet their use entails many pixel comparisons---a significant portion of which may contribute only marginally to the result~\citep{Cherel2024PSAL}---and stacking many layers to attain high-quality results.
This reliance on depth and scale limits interpretability and can also hinder deployment on devices with computation and memory constraints, where RAW denoising often takes place.
Although some efforts have been made to incorporate more classical notions of nonlocal self-similarity into denoising networks~\citep{Lefkimmiatis2017NLNet,Cruz2018Nonlocality,Yan2020Combining,Meng2024Nonlocal}, most of them embed classical priors within otherwise generic architectures rather than building new architectures around these notions.

In this work, we propose a deep neural network for RAW-to-RAW image denoising that explicitly translates the interpretable structure of classical nonlocal methods into an end-to-end learnable framework.
Our architecture is built around a novel Nonlocal Feature Matching and Filtering block that mirrors the three-stage pipeline of classical collaborative filtering methods.
Crucially, we operate on learned feature representations within a multiscale architecture rather than directly on image patches, allowing the network to leverage richer semantic information while maintaining the core principle of nonlocal self-similarity.
By also accepting a noise level map as input, the network adapts to spatially-varying noise, making it naturally suited for realistic camera noise models.
The network learns to identify matching neighboring features within a search window around each pixel for collaborative filtering.
This built-in nonlocality, combined with the multiscale structure, expands the receptive field efficiently without requiring the computational overhead of self-attention or excessive network depth.
The result is a denoising network that bridges classical methods with modern deep learning, retaining the theoretical grounding from the former while harnessing the representational power of the latter.

Our main contributions are as follows:
\begin{itemize}[nosep]
    \item We propose a novel, fully learnable neural module that performs nonlocal block matching, collaborative filtering and aggregation in an adapted feature domain, providing an alternative to black-box image denoising architectures.
    \item We model the noise profiles of a variety of camera sensors, using them to build a dataset with clean real RAW data and on-demand synthetic noise with which to train a network for sensor-agnostic RAW-to-RAW image denoising.
    \item We demonstrate that integrating the proposed block within a three-scale UNet archieves high-quality results that are comparable or superior to modern methods with deeply-stacked generic layers, all while using significantly fewer parameters.
    \item We validate our approach on the DND RAW benchmark~\citep{Plotz2017DND}, achieving state-of-the-art quantitative results, and also demonstrate strong qualitative performance on in-the-wild phone captures.
\end{itemize}


\section{Related Work}
\label{sec:relatedwork}

\subsection{Classical Self-similarity-based Denoising}

Image restoration has historically relied on mathematical models of the intrinsic characteristics of images and the relations that arise between their pixels.
Among classical methods, patch-based ones exploiting the self-similarity of images to denoise them have had the most long-lasting repercussion.
Nonlocal Means~\citep{Buades2005NLMeans} prompted this idea by proposing to eliminate noise by matching and averaging patches without explicitly encouraging their spatial proximity.
\citet{Kervrann2006Optimal} introduced spatial adaptation to it by refining patch weights based on a local estimation of their noise variance.
Deviating from kernel-based averaging, BM3D~\citep{Dabov2007BM3D} became the reference patch-based algorithm by using collaborative Wiener filtering in a fixed orthonormal basis.
The method popularized the general three-step architecture of patch matching, filtering and aggregation, which was subsequently used by algorithms like PLOW~\citep{Chatterjee2012PLOW}, NL-Bayes~\citep{Lebrun2013NLBayes}, and WNNM~\citep{Gu2014WNNM}, each employing different collaborative filtering strategies to exploit priors such as joint sparsity and low-rank signal representations.

\subsection{Denoising CNNs}

The arrival of CNNs caused a paradigm shift in the field of denoising.
DnCNN~\citep{Zhang2017DnCNN} showed that a simple feed-forward network composed of sequential convolutional layers, ReLU activations and batch normalization~\citep{Ioffe2015BatchNorm} could outperform classical methods considerably when trained end-to-end to learn the residual between clean and noisy images, despite working only locally.
The same year, IRCNN~\citep{Zhang2017IRCNN} used dilated convolutions to enlarge the receptive field, although at the cost of generating artifacts around sharp edges.
FFDNet~\citep{Zhang2018FFDNet} enlarged it in an alternative way: it applied DnCNN to four downsampled subimages obtained via pixel unshuffle~\citep{Shi2016PixelShuffle}, also concatenating a noise map to the input so as to help the network generalize to multiple noise levels.

Multiscale processing has since become widely adopted to aggregate distant information and alleviate the limitations imposed by local receptive fields.
SADNet~\citep{Chang2020SADNet} uses a spatial-adaptive block based on deformable convolutions~\citep{Dai2017Deformable} within a multiscale encoder-decoder architecture.
Through a neural architecture search, CLEARER~\citep{Gou2020CLEARER} learns when and how to extract and fuse cross-scale features.
Taking a different, interpretable approach, DeamNet~\citep{Ren2021DeamNet} unfolds a model-based denoiser with an adaptive consistency prior into a multiscale end-to-end trainable convolutional network.
MSANet~\citep{Gou2022MSANet} uses an asymmetric encoder-decoder architecture with different subnetworks per scale, fusing coarse details into finer scales with modulated deformable convolutions~\citep{Zhu2019DCNv2}.
Also taking a noise map as input, DRUNet~\citep{Zhang2022DPIR} achieves state-of-the-art results with a bias-free four-scale UNet~\citep{Ronneberger2015UNet} purely composed of residual blocks~\citep{He2016ResNet}, which the authors use in a Plug-and-Play scheme for different restoration tasks.

\subsection{Learning From Nonlocal Information}

The absence of explicit nonlocal mechanisms in CNNs motivated efforts to integrate the self-similarity principle into learning-based frameworks.
\citet{Lefkimmiatis2017NLNet} proposed applying block matching on noisy image patches and feed them to a CNN.
\citet{Qiao2017Nonlocal} also used block matching to embed a nonlocal self-similarity prior into the TNRD network~\citep{Chen2017TNRD}.
\citet{Liu2018NLRN} were the first to build a recurrent network with nonlocal operations, using soft matching on feature representations.
\citet{Plotz2018N3Net} designed a learnable block based on a differentiable relaxation of \(K\)-nearest neighbors selection, interleaving it with DnCNN modules.

Following a different direction from explicit patch matching, \citet{Wang2018NLNeural} proposed a general feed-forward block for nonlocal filtering, computing responses based on relationships between different locations.
This work regarded the self-attention mechanism used in language models~\citep{Vaswani2017Attention} as a form of nonlocal mean, leading to the development of nonlocal attention networks~\citep{Zhang2019RNAN,Mei2023PANet} and vision transformers~\citep{Dosovitskiy2021ViT}.
Since then, transformers have become state-of-the-art in image restoration~\citep{Chen2021IPT,Yin2022CSformer,Zhuge2023FeatureEnhanced,Li2024EWT,Zhou2024Efficient}.
In particular, starting with SwinIR~\citep{Liang2021SwinIR}, architectures based on shifted windowing schemes~\citep{Liu2021Swin} have had special success, reducing the computational cost of self-attention by limiting it to nonoverlapping local regions that are shifted after each layer.
Similar window-based transformer blocks have been used by networks like Uformer~\citep{Wang2022Uformer} and HWFormer~\citep{Tian2024HWFormer}: the first, in a UNet structure, and the second, separating shift directions in horizontal and vertical blocks.
As an alternative to window-based self-attention, Restormer~\citep{Zamir2022Restormer} proposes a transposed attention transformer block with a depthwise convolutional head, which is also embedded in a UNet-like architecture.
Works like CTNet~\citep{Tian2024CTNet}, Xformer~\citep{Zhang2024Xformer} and DSCA-Former~\citep{Hu2026DSCA} combine different transformer configurations in parallel to capture richer feature dependencies.

While early efforts to learn nonlocal filters employed explicit block matching, transformers have become the preferred alternative for denoising, as evidenced by their dominance in the latest benchmarks~\citep{NTIRE2025}.
This trend has left the former paradigm underexplored in recent literature, despite its principled connection to classical methods.
In this work, we show that this approach remains viable and competitive, building a network around the classical three-stage pipeline used by collaborative filtering algorithms.

\subsection{Denoising RAW images}

Not only the sensor noise model, but the distinct characteristics of RAW image data demand dedicated denoising strategies.
Unlike typical RGB images, RAW data has a mosaic structure---caused by a Color Filter Array (CFA)---in which spatial and chromatic information are interleaved.
Classical methods have been adapted to these CFA patterns by packing same-color pixels into separate channels and using color decorrelation transforms~\citep{Zhang2009CFA,Akiyama2015Pseudo,Buades2020CFA}, or by jointly denoising and demosaicking~\citep{Hirakawa2006JDD,Chatterjee2011NoiseSuppression,Tan2017ADMM}.
While Gaussian denoisers can be applied after variance stabilization to remove sensor noise~\citep{Makitalo2014NoiseMismatch}, the transformed noise distribution often exhibits heavier tails than a true Gaussian~\citep{Zhang2015Quantile}, resulting in inaccurate restoration under low-light conditions.

The emergence of deep learning has transformed the problem from one of algorithmic adaptation to one of data acquisition.
Several datasets have been introduced to meet this need.
To deal with especially problematic low signal-to-noise ratios in darker regions, works from~\citet{Chen2018LearningSeeDark}, \citet{Prabhakar2021Enhancement} and~\citet{Wei2022Physics} present datasets composed of low-light scenes accompanied with long-exposure clean images.
\citet{Brummer2025RawNIND} share a diverse collection of paired RAW images from a variety of camera models and CFA patterns.
For benchmarking, DND~\citep{Plotz2017DND} and SIDD~\citep{Abdelhamed2018SIDD} provide noisy images and a platform to evaluate denoised results against their held-out ground truths.
RAW denoising datasets have also been curated for the case of video, as is the case of CRVD~\citep{Yue2020RViDeNet} and its successor ReCRVD~\citep{Yue2025RViDeformer}.

Real data is often insufficient for effectively training RAW denoising networks, so synthesizing realistic noisy images becomes necessary.
Capitalizing on the abundance of noise-free sRGB image datasets, \citet{Brooks2019Unprocessing} invert the image processing pipeline to do so, and train a CNN on that unprocessed data.
Using a similar idea, CycleISP~\citep{Zamir2020CycleISP} learns simultaneously to generate synthetic RAW data from RGB images, and to denoise and process RAW images into RGB with a two-branch convolutional network.
Pseudo-ISP~\citep{Cao2024PseudoISP} follows the same direction, removing the need for RAW-RGB pairs to train the synthesis branch and generating pseudo-RAW images instead.
DualDn~\citep{Li2024DualDn} introduces a differentiable processing chain to perform dual-domain denoising in RAW and sRGB, adapting to both sensor noise and image signal processor variations.

\section{Method}

\begin{figure*}
    \centering
    \includegraphics[width=\linewidth]{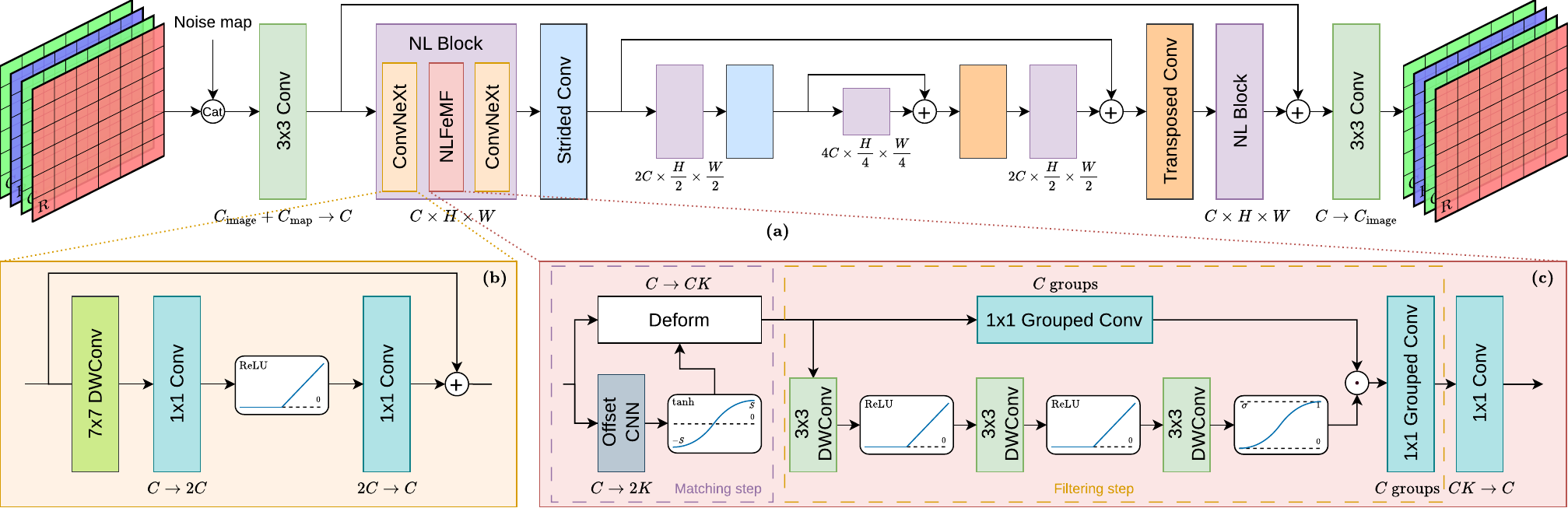}
    \caption{Architecture of the proposed denoising network. We use a three-scale UNet with a single NL block per scale. The NL block envelops a Nonlocal Feature Matching and Filtering block between two simplified ConvNeXt layers.}
    \label{fig:architecture}
\end{figure*}

We design our network for RAW-to-RAW denoising of images captured with a Bayer CFA.
To preserve their mosaic structure, we pack the RAW images into four \(RG_{r}BG_{b}\) channels, and then normalize them to the \([0, 1]\) range by removing their black level offset and dividing by their saturation value.
We feed the data to the network together with a noise map of its same size, concatenated along the channel dimension.
The map holds the standard deviation of the noise at each pixel, estimated from the shot and readout coefficients of the sensor assuming a regular affine variance model with respect to the intensity values.
Since each channel of the packed image corresponds to a distinct position, this noise map also has four channels;
this yields an eight-channel input to the network, which outputs a four-channel denoised packed RAW image that can be subsequently unpacked back.

\subsection{Overall Network Architecture}

Our proposed pipeline is inspired by the interpretable structure of classical nonlocal patch-based denoisers.
These methods have a common structure consisting in an initial block matching step, followed by collaborative filtering in a transformed domain, and a final aggregation of the filtered blocks---their main difference lying in the choice of transformation and the operator used to shrink the transform spectrum.
We maintain this basic three-step structure but deviate from the classical model-based approach, instead learning how to match image features to similar neighbors, linearly transform them, and find a suitable shrinkage operator to denoise them with a convolutional neural network.

The architecture of our proposed network is illustrated in Figure~\ref{fig:architecture}.
It adopts a three-scale UNet configuration, encoding image features by progressively expanding channel capacity while reducing the image dimensions, and decoding them in reverse to get a deep representation in the original resolution.
The reasoning behind this configuration is twofold.
Firstly, it increases the receptive field of the network, as downscaling the image allows the convolutional and nonlocal layers to reach into more distant features that are aggregated in upper scales.
Secondly, it mimics classical pyramidal approaches for noise removal~\citep{Burger2011Multiscale,Lebrun2015Multiscale,Facciolo2017Multiscale}, in which reconstructing the degraded image from the bottom up helps removing the lower frequencies of the noise.

A single convolutional layer is used to first extract shallow features from the input formed by the concatenated image and noise map, producing an initial noise-aware feature representation.
These shallow features are then passed to the UNet core.
In each scale, the features are processed with the same three sequential components: our proposed Nonlocal Feature Matching and Filtering (NLFeMF, detailed in Section~\ref{sec:nonlocalblock}), preceded and succeeded by ConvNeXt blocks~\citep{Liu2022ConvNeXt}, both without any normalization.
This arrangement (called NL block in Figure~\ref{fig:architecture}) serves a specific purpose: the ConvNeXt blocks, with their \(7 \times 7\) depthwise convolutions, provide essential local context extraction and feature refinement around each pixel, complementing the nonlocal processing performed by our feature matching and filtering block.

Downsampling is performed with \(2\)-strided \(2 \times 2\) convolutions, whereas we use their transposed equivalent for upsampling.
As is standard in UNets, skip connections link down and up blocks in corresponding scales, adding back the residual progressively while decoding.
Finally, a single convolutional layer decodes the refined high-resolution features into a denoised image of the same size and channels as the input noisy image.

\subsection{Nonlocal Feature Matching and Filtering Block}
\label{sec:nonlocalblock}

\begin{figure*}[t]
    \centering
    \includegraphics[width=0.98\linewidth]{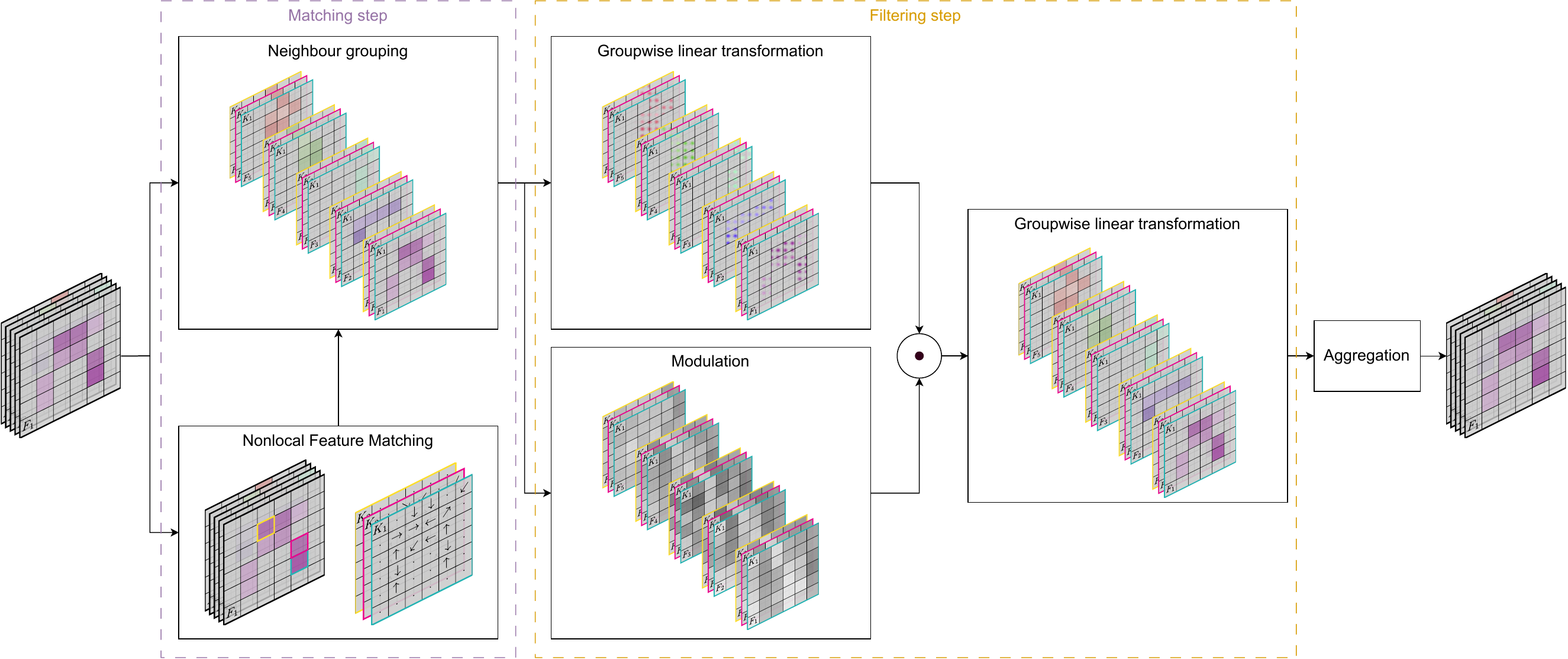}
    \caption{Outline of the proposed Nonlocal Feature Matching and Filtering block. All linear transformations, as well as the computation of the positional offsets for feature matching and of the modulation coefficients for collaborative filtering are learned.}
    \label{fig:nlfmblock}
\end{figure*}

The Nonlocal Feature Matching and Filtering (NLFeMF) block is the core component of our proposed network.
It implements a learnable variant of the block matching and collaborative filtering paradigm of classical patch-based methods, adapted to operate on learned feature representations rather than image patches.
A high-level illustration of its design is depicted in Figure~\ref{fig:nlfmblock}.
The block operates in three sequential steps: nonlocal feature matching, collaborative filtering, and aggregation.
Each of these steps is built of a specific configuration of convolution and deformation layers, designed to imitate the purpose of their classical counterparts.
Their detailed composition is pictured in the bottom right of Figure~\ref{fig:architecture}, and explained as follows.

\subsubsection{Feature Matching}

The first step identifies, for each pixel, the \(K\) optimal neighboring positions in a surrounding search window, based on their local features, to use afterward in collaborative filtering.
Rather than using a hand-crafted similarity metric, we learn the matching process directly.
Specifically, a convolutional neural network---akin to the offset estimation mechanism used by~\citet{Liang2022RVRT} in their guided deformable attention module for feature alignment between video frames---consisting of six \(3 \times 3\) convolutional layers with Leaky ReLU activations between them (with negative slope of \(0.1\)) predicts \(K\) two-dimensional offset vectors for each pixel.
A scaled hyperbolic tangent activation is used to restrict their values to the defined search window. 

The predicted offsets are not constrained to integer coordinates, so we apply bilinear interpolation to extract features at subpixel locations.
These neighboring features are then stacked on their corresponding pixel position, transforming a feature map with \(C\) channels into a grouped representation with \(C \cdot K\) channels (that is, each feature appears \(K\) times, one per neighbor).

\subsubsection{Collaborative Filtering}

To filter the stack of neighboring features, we first pass it through a \(1 \times 1\) grouped convolutional layer with \(C\) groups.
This operation learns a linear transformation \(T \colon \mathbb{R}^{K} \to \mathbb{R}^{K}\) that is applied independently to each of the \(C\) feature channels, effectively projecting the stacked neighbors into a representation adapted for denoising.
Note that grouping is essential so that there is no mixing of possibly-contrasting features, and that only closely resemblant matched values are transformed and filtered together.
This mirrors classical methods using a linear transform on a block of similar patches to obtain a sparse representation where signal and noise information can be easily separated.

\begin{figure}
    \centering
    \subcaptionbox{}{\includegraphics[width=0.64\linewidth]{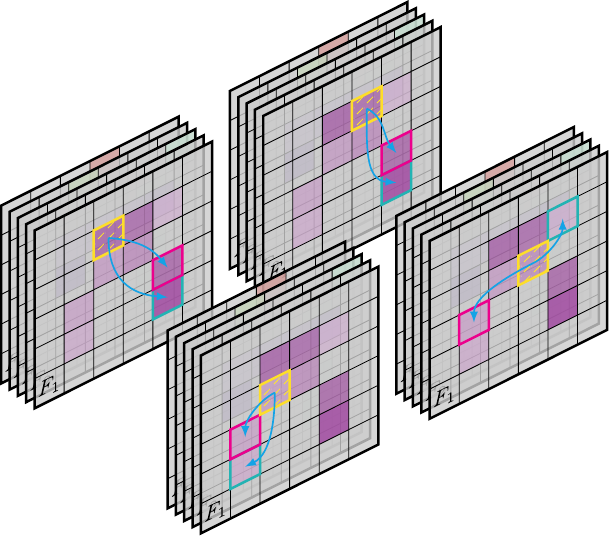}}%
    \hfill%
    \subcaptionbox{}{\includegraphics[width=0.31\linewidth]{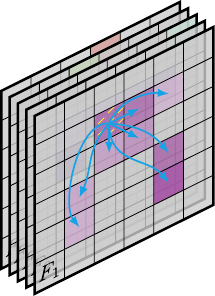}}%
    \caption{Local propagation of nonlocal neighboring features. (a) Convolving the stacked feature map aggregates nonlocal features in a local neighborhood, (b) effectively increasing the receptive field in a nonlocal manner.}
    \label{fig:featurepropagation}
\end{figure}

After transformation, we want to suppress the coefficients in the new representation that do not contribute to the image signal information and keep those that do.
We do so with a learned modulation map that shrinks the coefficients depending on the local structure of the stacked features.
This map is built by passing the whole non-transformed stack to a CNN composed of three \(3 \times 3\) depthwise convolutional layers with ReLU activations and a final sigmoid to get an attenuation coefficient between \(0\) and \(1\) for each one of the \(C \cdot K\) channels.
The depthwise convolutions operate spatially on the stacked features, allowing information from nonlocal neighbors at nearby pixel locations to interact.
As illustrated in Figure~\ref{fig:featurepropagation}, this effectively expands the receptive field by creating a larger nonlocal context.
We multiply the learned modulation map elementwise with the transformed feature stack to perform the filtering: coefficients which contain meaningful texture information (with a multiplier close to \(1\)) are preserved, those that do not (multiplier close to \(0\)) are suppressed, and intermediate values allow for partial attenuation, enabling the network to balance noise reduction with detail preservation in a differentiable manner.

Finally, we apply another transformation to reconstruct the feature stack from the shrunk coefficients.
Since the initial groupwise transformation \(T\) is not guaranteed to be invertible, we learn a separate mapping using another \(1 \times 1\) grouped convolutional layer with \(C\) groups, leading to another \(C \cdot K\)-channel representation of the stack of filtered features.

\subsubsection{Aggregation}

The final step reduces the \(CK\)-channel representation back to \(C\) channels.
Since the whole block of neighbors is filtered simultaneously for every position in the image, some of these neighbors may have multiple denoised estimates.
However, their features may have been interpolated at subpixel locations, so aggregating the neighbors back to their position with a simple averaging is an ill-defined operation.
Instead, we perform aggregation in feature space using a single \(1 \times 1\) convolutional layer that learns to optimally combine the filtered neighbors.

\subsection{RAW database}
\label{sec:adaptationraw}

Since the scarcity of high quality real data may hinder generalization to different sensors, we create our own synthetic noisy images for training.
However, we avoid unprocessing~\citep{Brooks2019Unprocessing}, as RAW data generated this way retains the irreversible quality loss of the original 8 bit images, which have already undergone quantization, tone mapping and compression.
Instead, we curate a dataset of high-quality ground truth RAW images from existing sources, to which we add synthetic Poisson-Gaussian noise.
We manually inspect each clean image in SID~\citep{Chen2018LearningSeeDark}, ELD~\citep{Wei2022Physics}, SIDD~\citep{Abdelhamed2018SIDD}, RawNIND~\citep{Brummer2025RawNIND}, Nikon~\citep{Prabhakar2021Enhancement}, CRVD~\citep{Yue2020RViDeNet} and ReCRVD~\citep{Yue2025RViDeformer}, discarding those with residual noise, and retain only Bayer CFA patterns.
For the video datasets, we select a single frame per scene to prevent bias from temporal redundancy.
This yields \(460\) images in the training set and \(30\) for validation.

\begin{figure}
    \centering
    \includegraphics[width=0.97\linewidth]{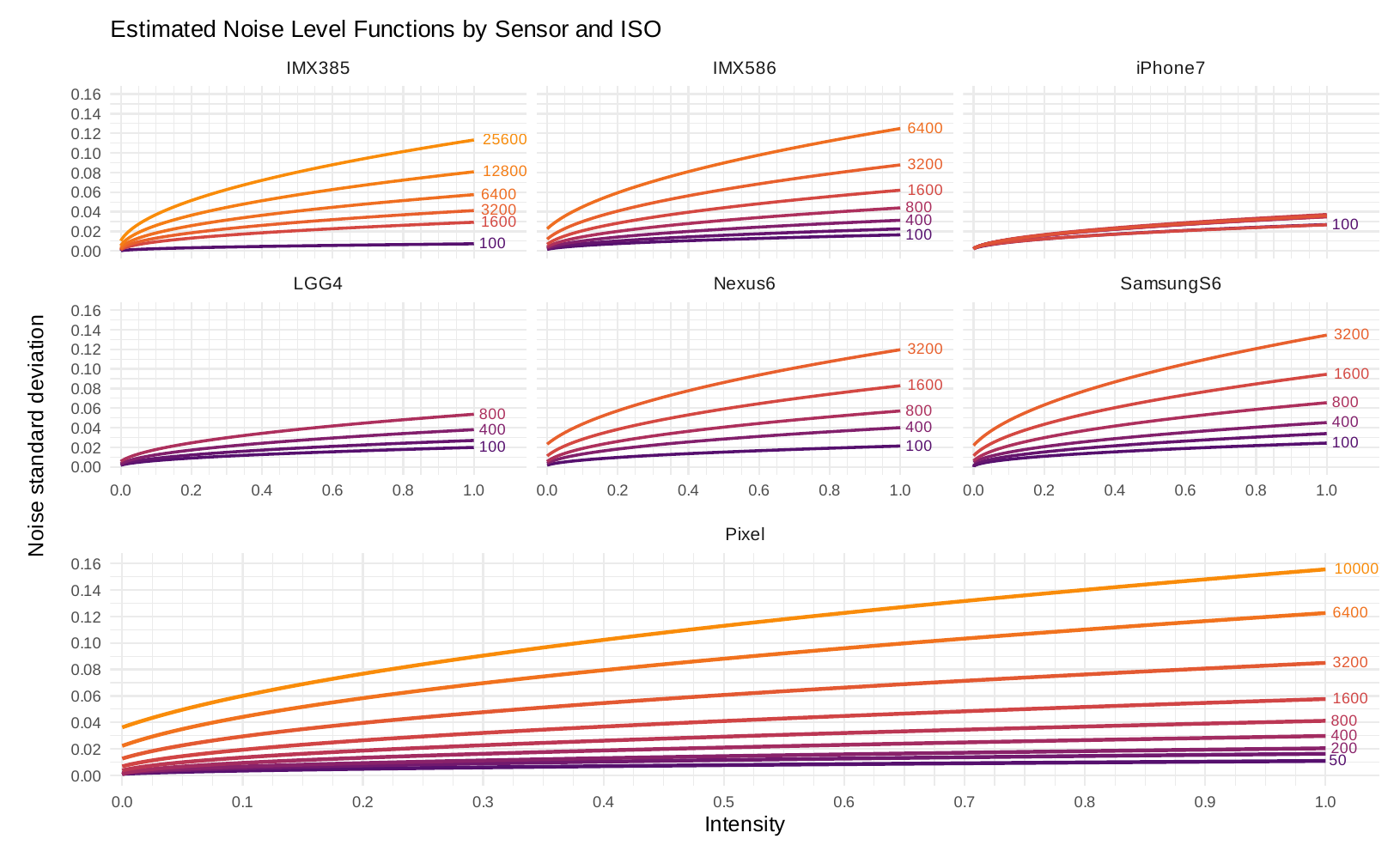}
    \caption{Estimated noise level curves of the used camera sensors for different ISO values (labeled to the right of the curves). The standard deviation of the noise is the square root of a linear model with respect to the pixel intensity. Data in the \([0, 1]\) range after black level subtraction.}
    \label{fig:noisecurves}
\end{figure}

To generate diverse enough noise, we sample the characteristics of seven different sensors at varying ISO levels.
We use the noise parameters of five sensors from the SIDD dataset (Samsung S6, iPhone 7, Google Pixel, Nexus 6, LG G4), as well as those from the Sony IMX385 and IMX586 sensors.
The parameters from the latter two are specified by~\citet{Yue2020RViDeNet} and~\citet{Wang2020Practical}, respectively.
Since the noise level functions embedded in SIDD have been shown to be miscalibrated~\citep{Zhang2021Rethinking}, we reestimate them from the data with the method proposed by~\citet{Colom2013Ponom}.
We measure variance levels for multiple intensity bins for all the images in the dataset, stratifying them by sensor model and ISO level, remove outliers found at high intensity values, and fit a linear curve to the measured points of each group.
Since the used method assumes that pixels in each intensity bin are corrupted by Gaussian noise, the slope \(a\) of the fitted curve and the intercept \(b\) are estimators of the shot and readout noise parameters in a signal-dependent heteroskedastic Gaussian noise model, i.e., the noise follows approximately a distribution \(\mathcal{N}(0, ax + b)\) for intensity \(x\).
The curves for the noise standard deviation of the used sensors are shown in Figure~\ref{fig:noisecurves}.
On noise generation during training, we use those same parameters with the more complex Poisson-Gaussian model.
We randomly select a sensor and an ISO value (within the available range for the sensor), interpolate the noise curve to that ISO, and generate the noisy data so that each noisy pixel is drawn as 
\begin{equation}
    x_{\text{noisy}} \sim a\mathcal{P}\left( \frac{x_{\text{true} }}{a} \right) + \mathcal{N}(0, b).
\end{equation}

This procedure yields diverse RAW training data that spans a realistic range of noise conditions while keeping the image quality encountered in real sensors.
Note that the noise standard deviation map during training must be estimated from the shot and readout coefficients using the noisy intensity values rather than the clean ones, so as to prevent the network from retrieving the ground truth signal from the noise map.

\subsection{Training details}

We train the network end-to-end with the \(L_{1}\) loss between denoised and clean images in the RAW domain.
On each training step, we crop a random square of size \(128 \times 128\) pixels from each one of the selected 4-channel packed images, augment it with a random transformation from the dihedral group \(D_{8}\), and add to it synthetic noise generated with the procedure explained in section~\ref{sec:adaptationraw}.
This is done for a total of \(10000\) epochs with a batch size of \(4\), amounting to 1.15M iterations.
We use Adam~\citep{Kingma2017Adam} as optimizer, starting with a learning rate of \(10^{{-4}}\) and gradually reducing it to \(5 \cdot 10^{{-7}}\) with a cosine annealing strategy~\citep{Loshchilov2017CosineAnnealing}.

\section{Network Analysis}

Before evaluating our method on real RAW data, we analyze the proposed architecture in the controlled setting of synthetic Gaussian noise.
This simplified scenario allows us to isolate the effects of individual design choices and compare different network configurations more easily.
We first describe the minor adaptations required to train the network for this setting, and then study how the number of neighbors, the multiscale structure, and the matching strategy affect denoising performance.
Finally, to position the proposed nonlocal architecture within the broader landscape of denoising paradigms, we perform quantitative and qualitative comparisons against distinguished classical, CNN-based and transformer-based Gaussian denoisers.

\subsection{Ablation study}

To adapt the proposed network to the Gaussian noise setting, we only modify the input and output channel sizes of its first and last layers, keeping the rest of the architecture as-is: the first layer receives \(4\)-channel data, where the last channel holds the same value of noise standard deviation for each pixel, while the last layer outputs \(3\)-channel RGB images.
Following previous works~\citep{Liang2021SwinIR,Zhang2022DPIR}, we train the network on images from the Waterloo Exploration Database~\citep{Ma2017WED}, DIV2K~\citep{Agustsson2017DIV2K}, and Flickr2K~\citep{Lim2017Flickr2K}, with grayscale images removed to avoid biases when generating color noise.
In total, after splitting the data, we use \(7969\) images for training and \(292\) for validation.
We add synthetic white Gaussian noise on demand at each step, choosing its standard deviation uniformly at random in the interval \([5, 50]\).
As the Gaussian noise dataset is larger than our RAW database, we decrease the number of epochs to \(2000\) and raise the batch size to \(16\), reaching around 1M iterations.
The chosen optimizer, learning rate, and scheduler are identical to the RAW setting.

All ablation experiments are performed on the CBSD68 test dataset~\citep{Martin2001CBSD68}, to which we add synthetic Gaussian noise with various standard deviations \(\sigma \in \{15,25,50\} \).
For quantitative comparisons, we use color PSNR as the evaluation metric.

\subsubsection{Effects of the number of neighbors}

\begin{table}
\centering
\caption{Effect of the number of neighbors \(K\). PSNR measured on CBSD68 for AWGN denoising. A slash / separates neighbors from different scales, from higher to lower resolution. Rows are grouped as: 1 scale, 3 scales.}%
\label{tab:numk}
\begin{tabular}[t]{ccccc}
\toprule
\multirow{2}{*}{\(K\)} & \multirow{2}{*}{Params (M)} & \multicolumn{3}{c}{PSNR (dB)} \\
 \cmidrule(l{3pt}r{3pt}){3-5}
 & & \(\sigma=15\) & \(\sigma=25\) & \(\sigma=50\) \\
\midrule
5  & 0.139          & 33.79 & 31.12 & 27.81 \\
9  & 0.322          & 33.90 & 31.23 & 27.95 \\
15 & 0.786          & 33.98 & 31.32 & 28.06 \\
25 & \phantom{0}2.1 & 34.03 & 31.37 & 28.11 \\
35 & \phantom{0}4.0 & 34.07 & 31.42 & 28.16 \\
49 & \phantom{0}7.7 & 34.11 & 31.45 & 28.18 \\
63 & 12.6           & 34.14 & 31.48 & 28.22 \\
81 & 20.8           & 34.15 & 31.50 & 28.24 \\
\midrule
5/5/5   & \phantom{0}3.5 & 34.23 & 31.61 & 28.41 \\ 
15/7/5  & \phantom{0}5.4 & 34.30 & 31.67 & 28.47 \\
15/9/7  & \phantom{0}7.5 & 34.31 & 31.69 & 28.50 \\
25/15/9 & 15.3           & 34.34 & 31.72 & 28.53 \\
\bottomrule
\end{tabular}
\end{table}

\begin{figure}
    \centering
    \captionsetup[subfigure]{labelformat=empty}
    \begin{minipage}{0.305\linewidth}%
        \subcaptionbox{Noisy}{%
            \includegraphics[width=\linewidth]{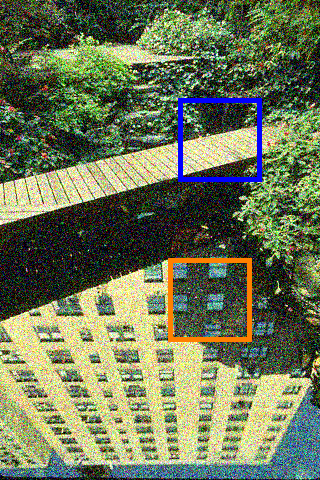}%
        }%
    \end{minipage}\hfill%
    \begin{minipage}{0.69\linewidth}%
        \includegraphics[width=0.31\linewidth,cframe=blue 1pt]{cbsd68-0026-15nbr-180\_100.png}\hfill%
        \includegraphics[width=0.31\linewidth,cframe=blue 1pt]{cbsd68-0026-49nbr-180\_100.png}\hfill%
        \includegraphics[width=0.31\linewidth,cframe=blue 1pt]{cbsd68-0026-15\_9\_7nbr-180\_100.png}%
        \\[2pt]%
        \subcaptionbox{15 nbr}{\includegraphics[width=0.31\linewidth,cframe=orange 1pt]{cbsd68-0026-15nbr-170\_260.png}}\hfill%
        \subcaptionbox{49 nbr}{\includegraphics[width=0.31\linewidth,cframe=orange 1pt]{cbsd68-0026-49nbr-170\_260.png}}\hfill%
        \subcaptionbox{15/9/7 nbr}{\includegraphics[width=0.31\linewidth,cframe=orange 1pt]{cbsd68-0026-15\_9\_7nbr-170\_260.png}}%
    \end{minipage}%
    \caption{Visual comparison between different values of \(K\) at different scales. A slash separates neighbors from different scales. Noisy image is corrupted with Gaussian noise with \(\sigma=50\).}%
    \label{fig:neighbors}
\end{figure}

To analyze the effect of varying the number of neighbors \(K\) used by the network, we build and train a simplified single-scale network containing an individual NL block plus the \(3 \times 3\) head and tail convolutions.
As shown in Table~\ref{tab:numk}, increasing \(K\) leads to higher PSNR values at the cost of additional network complexity.
The parameter increase is consequence of the filtering step needing to learn a set of linear filters for the stack of \(K\) matched neighbors.
The gains in PSNR diminish for large \(K\), as the additional neighbors become progressively more redundant.
These results go in accordance with classical patch-matching methods, where increasingly dissimilar patches contribute only marginally to the filtering.

Extending the architecture to multiple scales yields a substantial improvement in PSNR, as it can also be seen in Table~\ref{tab:numk}.
This improvement is noticeable even when the number of neighbors at each scale is low:
since spatial resolution decreases and feature dimensionality increases at coarser scales, fewer neighbors are required at those levels to maintain denoising quality.

Figure~\ref{fig:neighbors} shows a visual comparison between different arrangements, using a single scale with \(K=15\) and \(K=49\), and three scales with \(15\), \(9\) and \(7\) neighbors, respectively.
Looking at the blue patch, it is clear that increasing the number of neighbors on a single scale helps reconstruct structured patterns that repeat in the image.
Increasing the number of scales further improves the reconstruction quality, which is especially noticeable in fine textures like the lines on the bridge and the vegetation behind it.
Note that the latter two network configurations use approximately the same number of parameters (7.7M and 7.5M).
Even if each scale individually uses fewer neighbors than in the single scale case, the neighbors at lower scales retrieve spatial relations inside the image more effectively, which allows the network to discern noise from texture more precisely.
The use of multiple scales also helps remove low-frequency artifacts that appear at high noise levels, like the one in the orange patch: the hierarchical structure allows for a progressive and more robust neighbor selection, which translates to higher image quality.

\subsubsection{Feature matching strategy}

We compare our learned convolutional neighbor search against two alternatives: using a fixed local window around each pixel as the neighbor set, and the differentiable Patch Match algorithm proposed by~\citet{Cherel2024PSAL}.
We do so within a single scale, with \(K=15\) neighbors, or an equivalent \(3 \times 5\) window for the local baseline.

\begin{figure*}
    \centering
    \captionsetup[subfigure]{labelformat=empty}
    \begin{minipage}{0.195\linewidth}%
        \subcaptionbox{Noisy}{%
            \includegraphics[width=\linewidth]{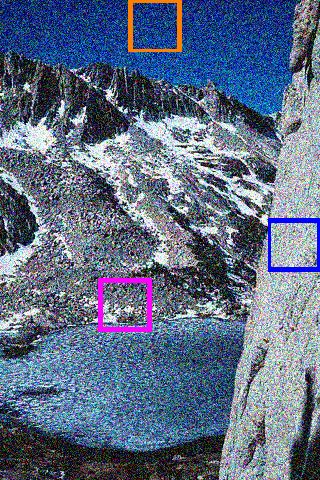}%
        }%
    \end{minipage}\hfill%
    \begin{minipage}{0.29\linewidth}%
        \includegraphics[width=0.31\linewidth,cframe=blue 1pt]{cbsd68-0033-15nbr\_cherel-270\_220.png}\hfill%
        \includegraphics[width=0.31\linewidth,cframe=blue 1pt]{cbsd68-0033-15nbr\_nooffset-270\_220.png}\hfill%
        \includegraphics[width=0.31\linewidth,cframe=blue 1pt]{cbsd68-0033-15nbr-270\_220.png}%
        \\[2pt]%
        \includegraphics[width=0.31\linewidth,cframe=orange 1pt]{cbsd68-0033-15nbr\_cherel-130\_0.png}\hfill%
        \includegraphics[width=0.31\linewidth,cframe=orange 1pt]{cbsd68-0033-15nbr\_nooffset-130\_0.png}\hfill%
        \includegraphics[width=0.31\linewidth,cframe=orange 1pt]{cbsd68-0033-15nbr-130\_0.png}%
        \\[2pt]%
        \subcaptionbox{Patch Match}{\includegraphics[width=0.31\linewidth,cframe=magenta 1pt]{cbsd68-0033-15nbr\_cherel-100\_280.png}}\hfill%
        \subcaptionbox{Local nbr}{\includegraphics[width=0.31\linewidth,cframe=magenta 1pt]{cbsd68-0033-15nbr\_nooffset-100\_280.png}}\hfill%
        \subcaptionbox{CNN offsets}{\includegraphics[width=0.31\linewidth,cframe=magenta 1pt]{cbsd68-0033-15nbr-100\_280.png}}%
    \end{minipage}%
    \hspace{0.02\linewidth}%
    \begin{minipage}{0.195\linewidth}%
        \subcaptionbox{Noisy}{%
            \includegraphics[width=\linewidth]{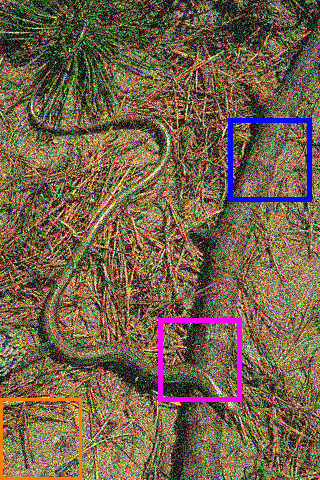}%
        }%
    \end{minipage}\hfill%
    \begin{minipage}{0.29\linewidth}%
        \includegraphics[width=0.31\linewidth,cframe=blue 1pt]{cbsd68-0035-15nbr\_cherel-230\_120.png}\hfill%
        \includegraphics[width=0.31\linewidth,cframe=blue 1pt]{cbsd68-0035-15nbr\_nooffset-230\_120.png}\hfill%
        \includegraphics[width=0.31\linewidth,cframe=blue 1pt]{cbsd68-0035-15nbr-230\_120.png}%
        \\[2pt]%
        \includegraphics[width=0.31\linewidth,cframe=orange 1pt]{cbsd68-0035-15nbr\_cherel-0\_400.png}\hfill%
        \includegraphics[width=0.31\linewidth,cframe=orange 1pt]{cbsd68-0035-15nbr\_nooffset-0\_400.png}\hfill%
        \includegraphics[width=0.31\linewidth,cframe=orange 1pt]{cbsd68-0035-15nbr-0\_400.png}%
        \\[2pt]%
        \subcaptionbox{Patch Match}{\includegraphics[width=0.31\linewidth,cframe=magenta 1pt]{cbsd68-0035-15nbr\_cherel-160\_320.png}}\hfill%
        \subcaptionbox{Local nbr}{\includegraphics[width=0.31\linewidth,cframe=magenta 1pt]{cbsd68-0035-15nbr\_nooffset-160\_320.png}}\hfill%
        \subcaptionbox{CNN Offsets}{\includegraphics[width=0.31\linewidth,cframe=magenta 1pt]{cbsd68-0035-15nbr-160\_320.png}}%
    \end{minipage}\hfill%
    \caption{Visual comparison between different feature matching strategies on two images from the CBSD68 dataset. A single scale with a NL block with \(K=15\) is used for denoising. Noisy images are corrupted with Gaussian noise with \(\sigma=50\).}%
    \label{fig:matching}
\end{figure*}

\begin{table}
\centering
\caption{Effect of the feature matching strategy. A single scale with a NL block with \(K=15\) is used. PSNR measured on CBSD68 for AWGN denoising.}%
\label{tab:matching}
\begin{tabular}[t]{lccc}
\toprule
\multirow{2}{*}{Matching strategy} & \multicolumn{3}{c}{PSNR (dB)} \\
 \cmidrule(l{3pt}r{3pt}){2-4}
 & \(\sigma=15\) & \(\sigma=25\) & \(\sigma=50\) \\
\midrule
Differentiable Patch Match & 33.76 & 31.09 & 27.63 \\
Local neighbors  & 33.92 & 31.24 & 27.93 \\
Offset CNN       & 33.98 & 31.32 & 28.06 \\
\bottomrule
\end{tabular}
\end{table}

Table~\ref{tab:matching} shows that the learned CNN-based feature matching outperforms the local window baseline, confirming the benefit of adaptively selecting neighbors.
The differentiable Patch Match performs notably worse, particularly at high noise levels.
This is illustrated in Figure~\ref{fig:matching}.
While Patch Match does a good job reconstructing textured areas like the wall and the log in the blue patches, it produces low-frequency artifacts in flat regions (orange patches).
Results using this matching strategy can also suffer from color bleed.
This can be seen in the magenta patches, where the gray rocks on the mountain have a blue tint, and the green snake adopts the browner color of the log.
We attribute these limitations to the difficulty of matching noisy features in a high-dimensional space, and the consequent profileration of incorrectly computed shifts during the propagation step of the algorithm.
Compared to the local baseline, the CNN-based matching leads to better reconstruction of texture in all cases.

\subsection{Comparison with Gaussian denoisers}

For completeness and to facilitate comparison with the broader denoising literature, we measure our method against the state of the art in AWGN denoising.
We compare it against denoisers spanning classical, CNN-based and transformer-based paradigms: BM3D~\citep{Dabov2007BM3D} for the former; DnCNN~\citep{Zhang2017DnCNN}, FFDNet~\citep{Zhang2018FFDNet} and DRUNet~\citep{Zhang2022DPIR} for the second; and Restormer~\citep{Zamir2022Restormer}, CTNet~\citep{Tian2024CTNet} and DSCA-Former~\citep{Hu2026DSCA} for the latter.
For all methods, we use their officially released parameter checkpoints trained on the broadest available noise range.
Quantitative results are reported on CBSD68~\citep{Martin2001CBSD68}, Kodak~\citep{Franzen1999Kodak} and McMaster~\citep{Zhang2011McMaster} for \(\sigma \in \{15, 25, 50\}\).

\begin{table*}
\centering
\caption{Average PSNR (dB) of various methods on test datasets for AWGN denoising. All compared methods use the same set of parameter weights across all noise standard deviations. Methods are grouped as: classical, CNN-based, transformer-based, and ours.}%
\label{tab:sotaawgn}
\begin{tabular}[t]{lcccccccccc}
\toprule
\multirow{2}{*}{Method} & \multirow{2}{*}{Params (M)} & \multicolumn{3}{c}{\(\sigma=15\)} & \multicolumn{3}{c}{\(\sigma=25\)} & \multicolumn{3}{c}{\(\sigma=50\)} \\
 \cmidrule(l{3pt}r{3pt}){3-5} \cmidrule(l{3pt}r{3pt}){6-8} \cmidrule(l{3pt}r{3pt}){9-11}
 & & CBSD68 & Kodak & McM & CBSD68 & Kodak & McM & CBSD68 & Kodak & McM \\
\midrule
BM3D           & - & 33.52 & 34.28 & 34.06 & 30.71 & 32.15 & 31.66 & 27.38 & 28.46 & 28.51 \\
\midrule
DnCNN          & 0.854 & 33.90 & 34.60 & 33.45 & 31.24 & 32.14 & 31.52 & 27.95 & 28.95 & 28.62 \\
FFDNet         & 0.854 & 33.87 & 34.63 & 34.66 & 31.21 & 32.13 & 32.35 & 27.96 & 28.98 & 29.18 \\
DRUNet         & 32.6 & 34.30 & 35.31 & 35.40 & 31.69 & 32.89 & 33.14 & 28.51 & 29.86 & 30.08 \\
\midrule 
Restormer      & 26.1 & 34.39 & 35.44 & 35.55 & 31.78 & 33.02 & 33.31 & 28.59 & 30.00 & 30.29 \\
CTNet          & 49.0 & 34.32 & 35.24 & 35.46 & 31.68 & 32.79 & 33.17 & 28.41 & 29.65 & 30.00 \\
DSCA-Former    & 65.1 & 34.35 & 35.42 & 35.58 & 31.72 & 32.87 & 33.28 & 28.48 & 29.67 & 30.21 \\
\midrule 
Ours (15 nbr)      & 0.786          & 33.98 & 34.88 & 34.84 & 31.32 & 32.38 & 32.52 & 28.06 & 29.21 & 29.31 \\
Ours (15/7/5 nbr)  & \phantom{0}5.4 & 34.30 & 35.30 & 35.37 & 31.67 & 32.87 & 33.11 & 28.47 & 29.82 & 30.02 \\
Ours (15/9/7 nbr)  & \phantom{0}7.5 & 34.31 & 35.33 & 35.40 & 31.69 & 32.90 & 33.14 & 28.50 & 29.85 & 30.07 \\
Ours (25/15/9 nbr) & 15.3           & 34.34 & 35.37 & 35.45 & 31.72 & 32.94 & 33.19 & 28.53 & 29.90 & 30.12 \\
\bottomrule
\end{tabular}
\end{table*}

Table~\ref{tab:sotaawgn} summarizes the obtained PSNR values.
Our three-scale configuration with 15/9/7 neighbors matches or exceeds DRUNet on all datasets, despite using less than one-fourth of the parameters (7.5M vs.\ 32.6M).
Using fewer neighbors in the coarsest scales to a 15/7/5 configuration reduces the number of parameters even further (5.4M) while maintaining competitive results.
The larger 25/15/9 variant performs comparably to CTNet and DSCA-Former at low noise levels and surpasses them at \(\sigma=50\), despite having significantly fewer parameters (15.3M vs.\ 49M and 65.1M).
It also offers lower network complexity than Restormer (26.1M parameters), against which the gap remains small.
On the opposite end of complexity, the lightweight single-scale version of our network (786k parameters) consistently outperforms BM3D, DnCNN and FFDNet, demonstrating that the proposed nonlocal block can be suitable by itself for low-resource environments.

\begin{figure*}
    \centering
    \captionsetup[subfigure]{labelformat=empty}
    \begin{minipage}{0.326\linewidth}%
        \includegraphics[width=\linewidth]{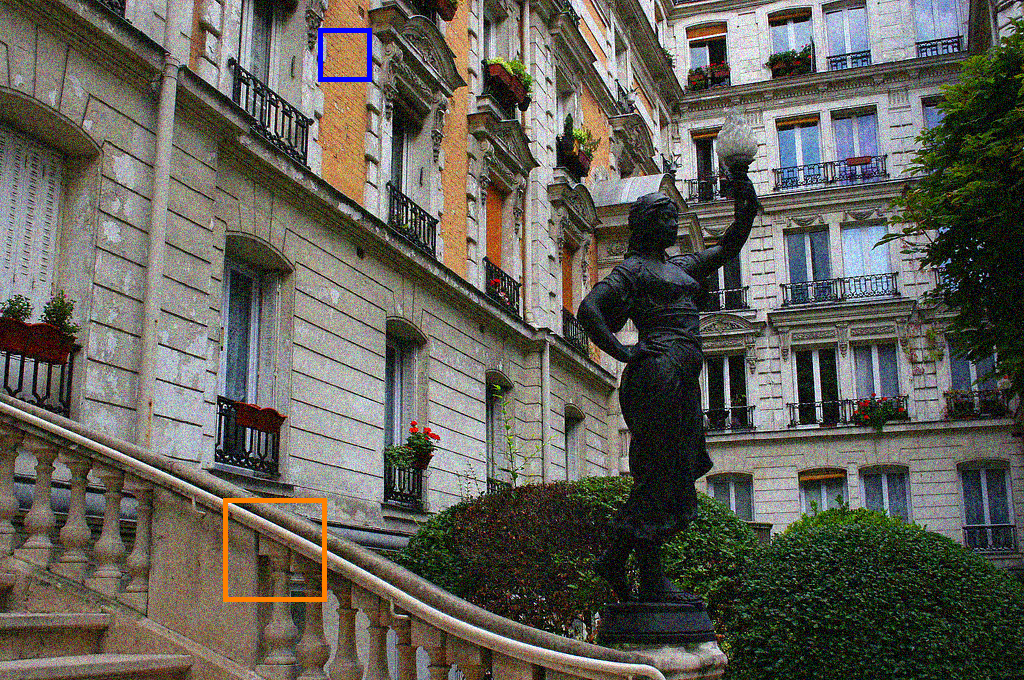}%
        \\[3pt]%
        \subcaptionbox{Noisy full image}{%
            \includegraphics[width=\linewidth]{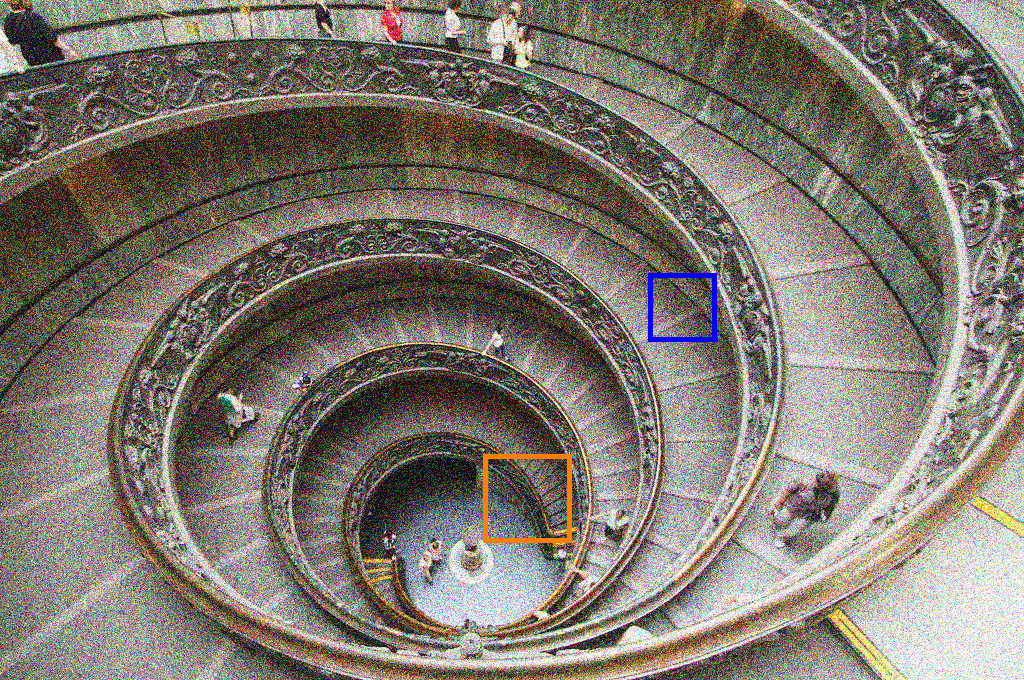}%
        }%
    \end{minipage}\hfill%
    \begin{minipage}{0.664\linewidth}%
        \includegraphics[width=0.155\linewidth,cframe=blue 1pt]{urban100-0002-gt-320\_030.png}\hfill%
        \includegraphics[width=0.155\linewidth,cframe=blue 1pt]{urban100-0002-noisy25-320\_030.png}\hfill%
        \includegraphics[width=0.155\linewidth,cframe=blue 1pt]{urban100-0002-drunet-320\_030.png}\hfill%
        \includegraphics[width=0.155\linewidth,cframe=blue 1pt]{urban100-0002-restormer-320\_030.png}\hfill%
        \includegraphics[width=0.155\linewidth,cframe=blue 1pt]{urban100-0002-ctnet-320\_030.png}\hfill%
        \includegraphics[width=0.155\linewidth,cframe=blue 1pt]{urban100-0002-25\_15\_9nbr-320\_030.png}%
        \\[0.5pt]%
        \includegraphics[width=0.155\linewidth,cframe=orange 1pt]{urban100-0002-gt-225\_500.png}\hfill%
        \includegraphics[width=0.155\linewidth,cframe=orange 1pt]{urban100-0002-noisy25-225\_500.png}\hfill%
        \includegraphics[width=0.155\linewidth,cframe=orange 1pt]{urban100-0002-drunet-225\_500.png}\hfill%
        \includegraphics[width=0.155\linewidth,cframe=orange 1pt]{urban100-0002-restormer-225\_500.png}\hfill%
        \includegraphics[width=0.155\linewidth,cframe=orange 1pt]{urban100-0002-ctnet-225\_500.png}\hfill%
        \includegraphics[width=0.155\linewidth,cframe=orange 1pt]{urban100-0002-25\_15\_9nbr-225\_500.png}%
        \\[3pt]%
        \includegraphics[width=0.155\linewidth,cframe=blue 1pt]{urban100-0050-gt-650\_275.png}\hfill%
        \includegraphics[width=0.155\linewidth,cframe=blue 1pt]{urban100-0050-noisy50-650\_275.png}\hfill%
        \includegraphics[width=0.155\linewidth,cframe=blue 1pt]{urban100-0050-drunet-650\_275.png}\hfill%
        \includegraphics[width=0.155\linewidth,cframe=blue 1pt]{urban100-0050-restormer-650\_275.png}\hfill%
        \includegraphics[width=0.155\linewidth,cframe=blue 1pt]{urban100-0050-ctnet-650\_275.png}\hfill%
        \includegraphics[width=0.155\linewidth,cframe=blue 1pt]{urban100-0050-25\_15\_9nbr-650\_275.png}%
        \\[0.5pt]%
        \subcaptionbox{Ground truth}{\includegraphics[width=0.155\linewidth,cframe=orange 1pt]{urban100-0050-gt-485\_456.png}}\hfill%
        \subcaptionbox{Noisy}{\includegraphics[width=0.155\linewidth,cframe=orange 1pt]{urban100-0050-noisy50-485\_456.png}}\hfill%
        \subcaptionbox{DRUNet}{\includegraphics[width=0.155\linewidth,cframe=orange 1pt]{urban100-0050-drunet-485\_456.png}}\hfill%
        \subcaptionbox{Restormer}{\includegraphics[width=0.155\linewidth,cframe=orange 1pt]{urban100-0050-restormer-485\_456.png}}\hfill%
        \subcaptionbox{CTNet}{\includegraphics[width=0.155\linewidth,cframe=orange 1pt]{urban100-0050-ctnet-485\_456.png}}\hfill%
        \subcaptionbox{Ours}{\includegraphics[width=0.155\linewidth,cframe=orange 1pt]{urban100-0050-25\_15\_9nbr-485\_456.png}}%
    \end{minipage}%
    \caption{Visual comparison between DRUNet, Restormer, CTNet and our method with 25/15/9 neighbors. The top image is corrupted with Gaussian noise with \(\sigma=25\), while the bottom one has \(\sigma=50\).}%
    \label{fig:awgn}
\end{figure*}

Figure~\ref{fig:awgn} shows a visual example comparing DRUNet, Restormer, CTNet and our 25/15/9 variant on images of the Urban100 dataset~\citep{Huang2015Urban100}.
All methods yield high quality results, with only subtle differences.
CTNet and our method produce slightly better reconstructions of thin straight lines, as it can be seen in the center and bottom right of the top blue patch, where the other two methods fade them too much.
Restormer tends to smooth out flat regions with low-frequency grain texture, like the shaded region at the left of the balustrade in the top orange patch.
The other three methods do not exhibit this issue as notably.
The patches of the bottom image highlight how both Restormer and CTNet can generate hallucinated textures when trying to reconstruct fine details.
In the blue patch, Restormer erroneously extends the step of the staircase into the wall, creating a blending effect.
DRUNet and our method correctly filter the texture on the wall; CTNet manages to further recover some of the stonelike pattern on it, but causes a ringing artifact around the step.
CTNet also fabricates a curved line between steps of the staircase in the orange patch, which clearly should not be present.
Overall, our method achieves a visually pleasant reconstruction that completely eliminates noise while precisely recovering fine details and avoiding hallucinations, all with a reduced parameter budget.

\section{Experimental Results on Real RAW Data}

We finally conduct experiments to assess the performance of our method on real RAW data.
We evaluate it on two complementary bases: the Darmstadt Noise Dataset (DND)~\citep{Plotz2017DND} for standardized testing, and the in-the-wild smartphone captures provided by~\citet{Li2024DualDn} for visual assessment of generalization across different sensors.

\subsection{Results on DND}

\begin{table}
\centering
\caption{Results on the Darmstadt Noise Dataset benchmark for RAW image denoising. The values are measured in the sRGB domain after applying the benchmark's own processing pipeline.}%
\label{tab:dnd}
\begin{tabular}[t]{lcccc}
\toprule
Method & PSNR (dB) & SSIM \\
\midrule
UPI       & 40.35 & 0.9641 \\
CycleISP  & 40.50 & 0.9655 \\
PseudoISP & 40.36 & 0.9606 \\
DualDn    & 40.70 & 0.9635 \\
Ours  (25/15/9 nbr)     & 40.63 & 0.9644 \\
\bottomrule
\end{tabular}
\end{table}

\begin{figure*}
    \centering
    \captionsetup[subfigure]{labelformat=empty}
    \begin{minipage}{0.215\linewidth}
    \includegraphics[width=0.99\linewidth]{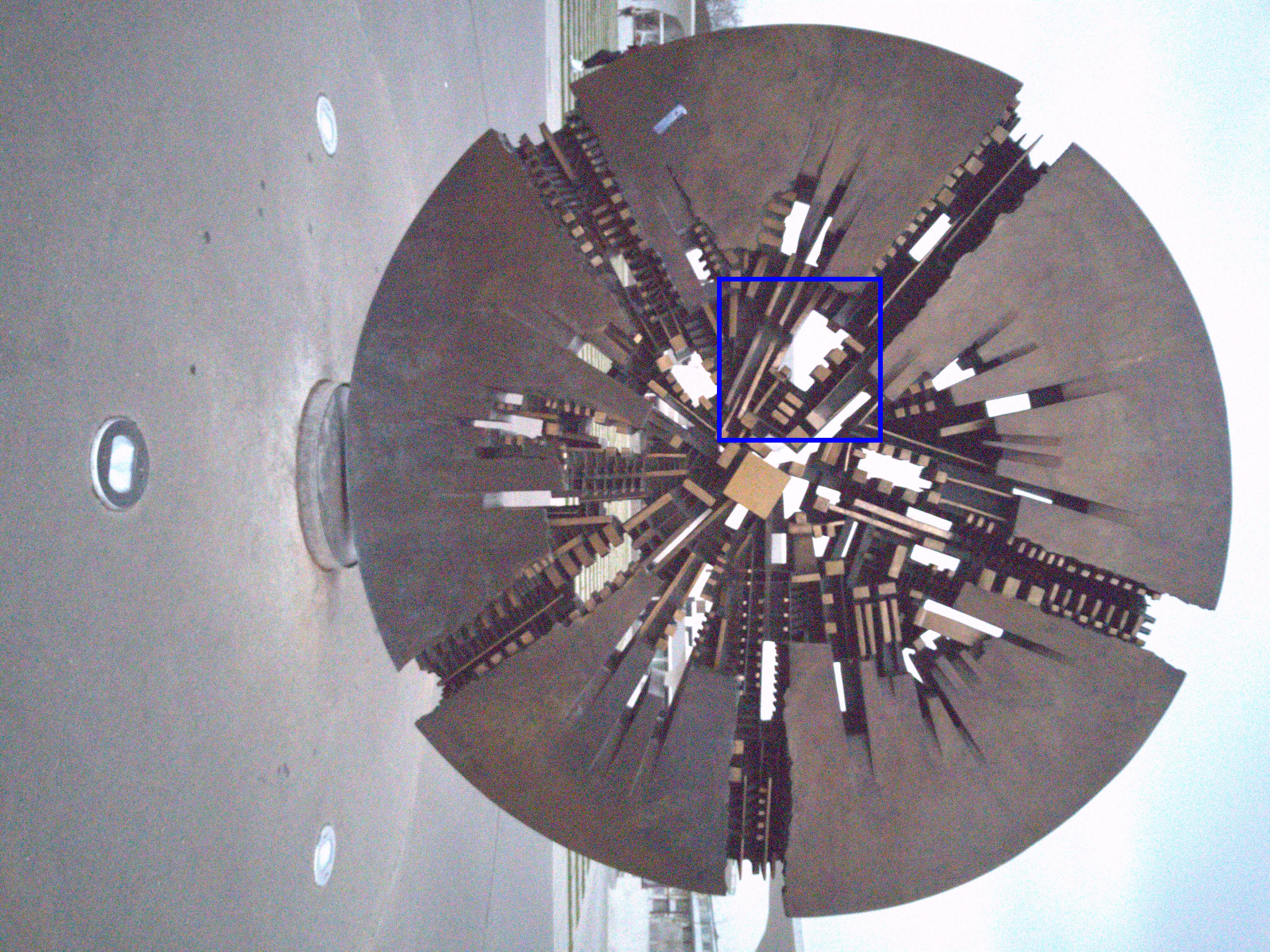}%
    \\[2pt]%
    \subcaptionbox{Noisy full image}{\includegraphics[width=0.99\linewidth]{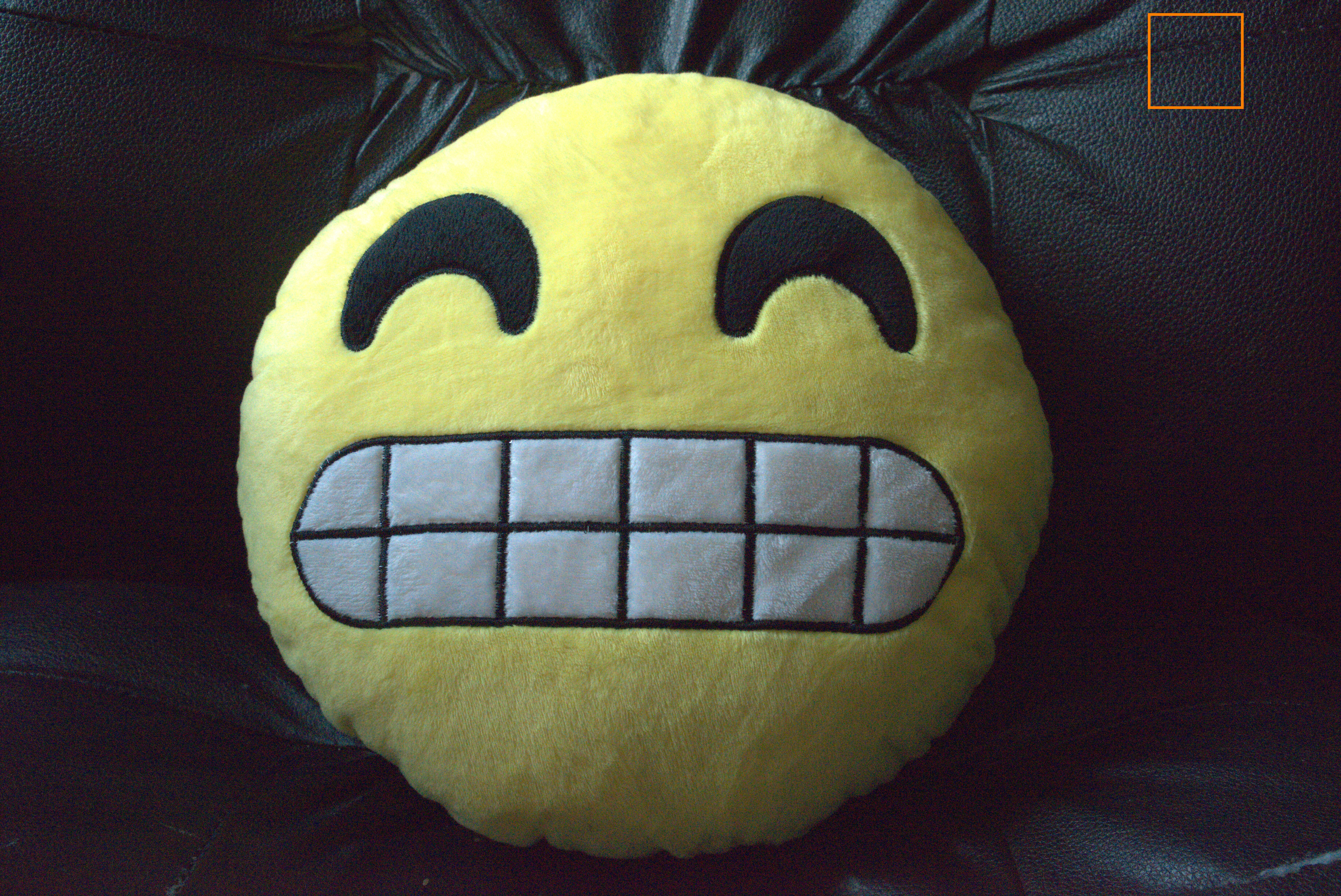}}%
    \end{minipage}\hfill%
    \begin{minipage}{0.775\linewidth}
    \includegraphics[width=0.19\linewidth,cframe=blue 1pt]{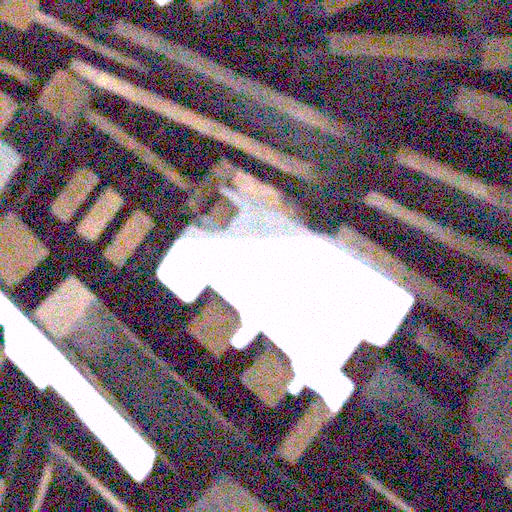}\hfill%
    \includegraphics[width=0.19\linewidth,cframe=blue 1pt]{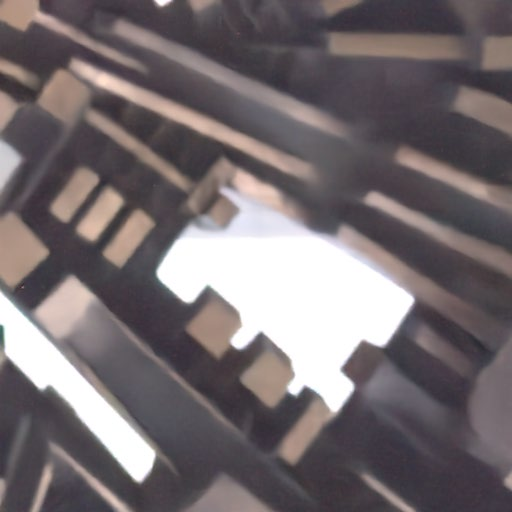}\hfill%
    \includegraphics[width=0.19\linewidth,cframe=blue 1pt]{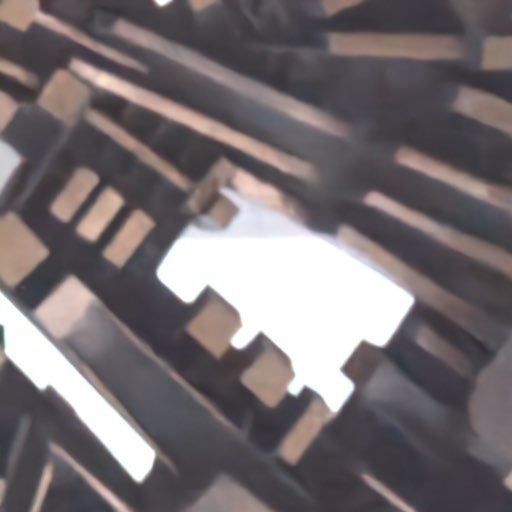}\hfill%
    \includegraphics[width=0.19\linewidth,cframe=blue 1pt]{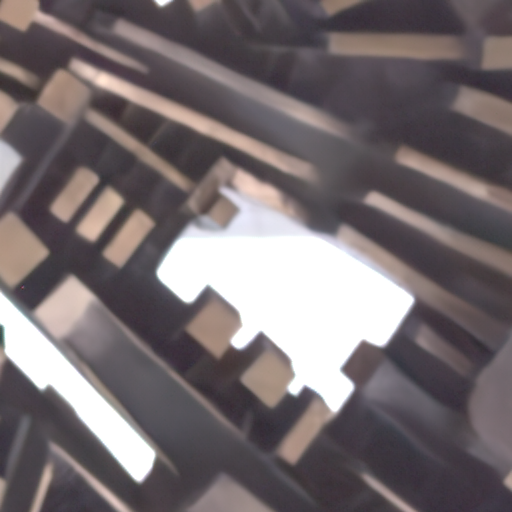}\hfill%
    \includegraphics[width=0.19\linewidth,cframe=blue 1pt]{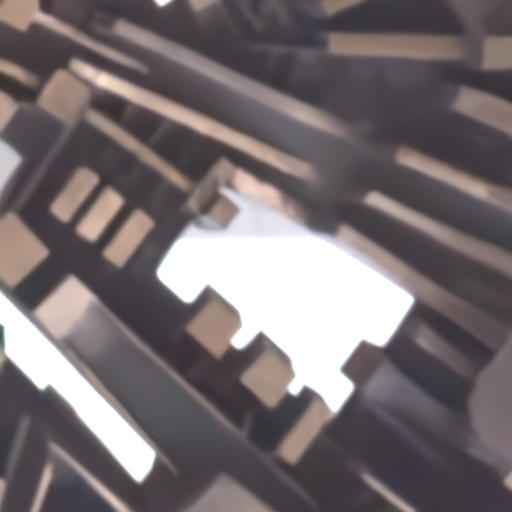}%
    \\[2pt]%
    \subcaptionbox{Noisy}{\includegraphics[width=0.19\linewidth,cframe=orange 1pt]{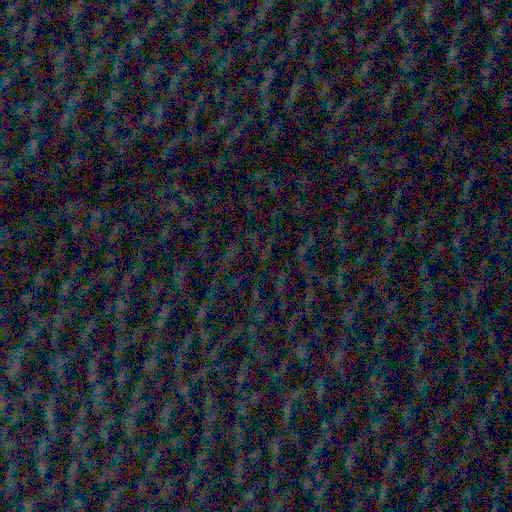}}\hfill%
    \subcaptionbox{UPI}{\includegraphics[width=0.19\linewidth,cframe=orange 1pt]{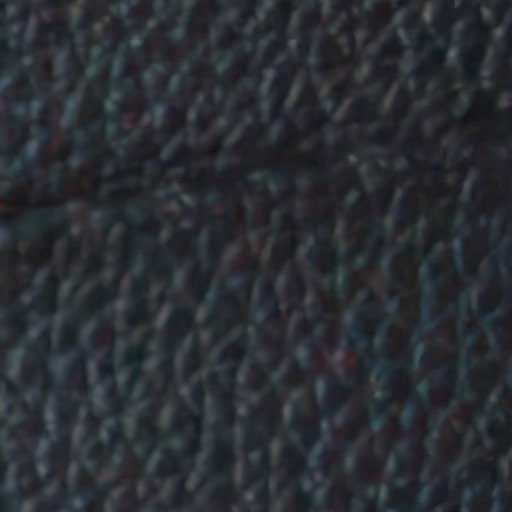}}\hfill%
    \subcaptionbox{CycleISP}{\includegraphics[width=0.19\linewidth,cframe=orange 1pt]{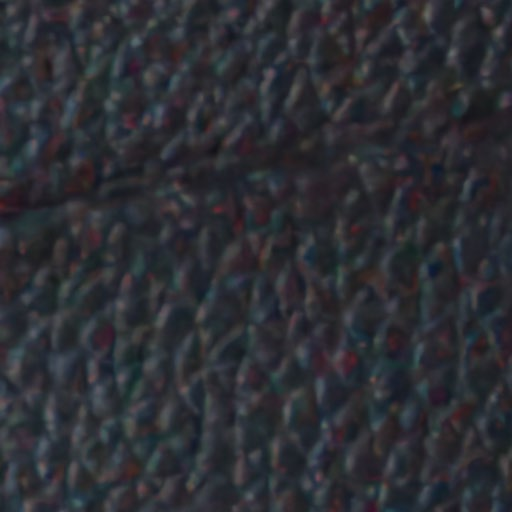}}\hfill%
    \subcaptionbox{DualDn}{\includegraphics[width=0.19\linewidth,cframe=orange 1pt]{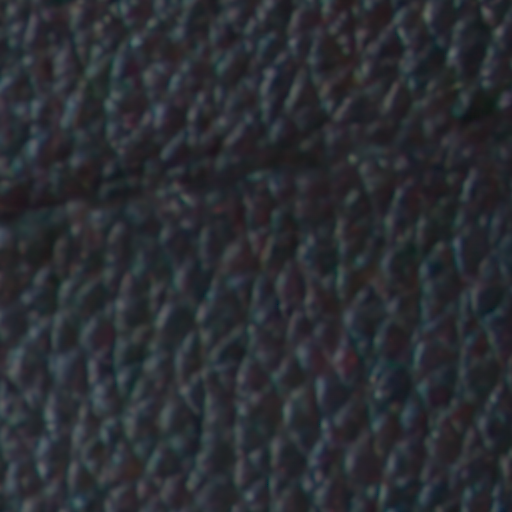}}\hfill%
    \subcaptionbox{Ours}{\includegraphics[width=0.19\linewidth,cframe=orange 1pt]{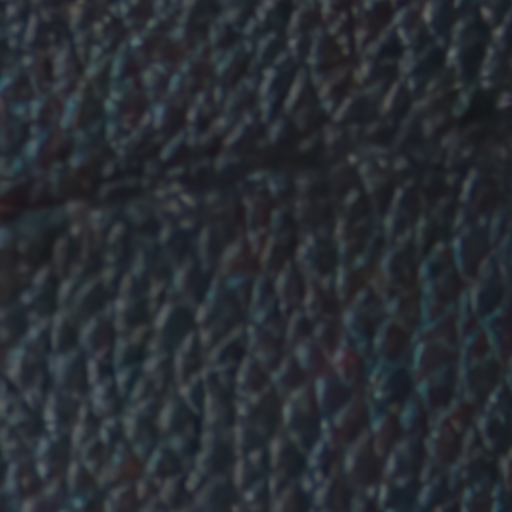}}%
    \end{minipage}%
    \caption{Visual comparison against state-of-the-art RAW denoising methods on the DND dataset. Denoised images are retrieved from the official benchmarking platform after being processed with an internal pipeline.}%
    \label{fig:dnd}
\end{figure*}

We compare our 25/15/9 network variant against the top-performing methods on the RAW denoising benchmark shared in the DND web platform: UPI~\citep{Brooks2019Unprocessing}, CycleISP~\citep{Zamir2020CycleISP}, PseudoISP~\citep{Cao2024PseudoISP}, and DualDn~\citep{Li2024DualDn}.
Note that DualDn is a modular RAW-to-RGB scheme that denoises both before and after demosaicking, and its reported metrics are those of the first RAW-to-RAW module, which uses Restormer~\citep{Zamir2022Restormer} as a backbone.
Each of the \(50\) test images of the dataset are assigned \(20\) bounding boxes that delimit the regions that are used for benchmarking; we denoise each of these crops individually, using the noise parameters embedded in the full image metadata to generate their noise maps.

Table~\ref{tab:dnd} reports the average PSNR and SSIM in the sRGB domain after applying the benchmark's processing pipeline on the denoised images.
Our method surpasses UPI, CycleISP and PseudoISP, showing a notable PSNR advantage.
DualDn achieves the highest PSNR---consistent with Restormer's results in the AWGN setting---but shows some regression in terms of SSIM, where we attain a higher value.
Visual inspection of the two crops in Figure~\ref{fig:dnd} (extracted in sRGB from the web platform) reveal some of DualDn's limitations that may partially explain the metrics' discrepancies.
Residual noise persists in darker flat regions, being especially noticeable at the bottom right of the blue patch.
This problem is also present in UPI, where the grain is even more discernible.
Also in that same patch, a hot-pixel artifact remains visible above the leftmost white region in both UPI and DualDn.
CycleISP does not leave residual noise and corrects the artifact, but has trouble reconstructing straight edges, bending them slightly.
Our method completely eliminates noise and the artifact, and produces a visually pleasant reconstruction.
Images from PseudoISP are unavailable.
In the orange patch, it can also be seen that all denoisers (especially CycleISP) exhibit a slight color bias toward red in what should be a uniformly black sofa.
This is low-frequency residual noise that is common in dark regions with low signal-to-noise ratio.
Our method substantially attenuates this bias, yielding more visually consistent results.

\subsection{Results on in-the-wild images}

The images in DND have been captured in a controlled setting, where most of them display bright, daylight scenes and are corrupted with a relatively low noise level. 
As consequence, results in the benchmark may not be representative of a method's capability of denoising daily life photographs.
To assess generalization to a bigger variety of sensors and image conditions, we perform a visual evaluation on the in-the-wild dataset collected by~\citet{Li2024DualDn}.
This dataset comprises high-resolution noisy RAW images captured by smartphone cameras of three different brands with a diversity of ISO values, of which a ground truth is not available.

\begin{figure*}
    \centering
    \captionsetup[subfigure]{labelformat=empty}
    \begin{minipage}{0.325\linewidth}
    \subcaptionsetup[figure]{position=top}
    \subcaptionbox{Huawei (ISO 1600)}{\includegraphics[width=\linewidth]{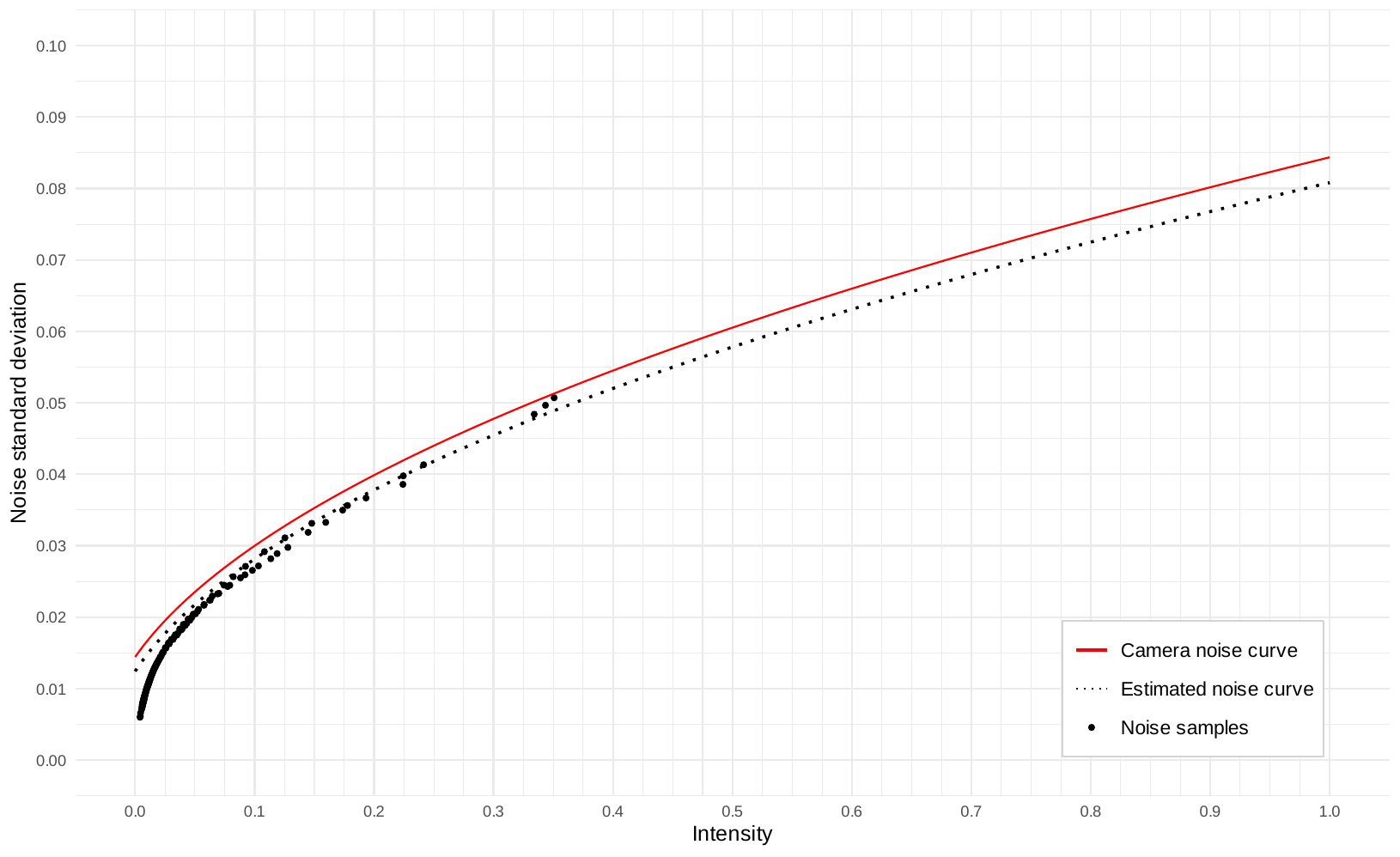}}%
    \\[2pt]%
    \subcaptionsetup[figure]{position=bottom}
    \includegraphics[width=\linewidth]{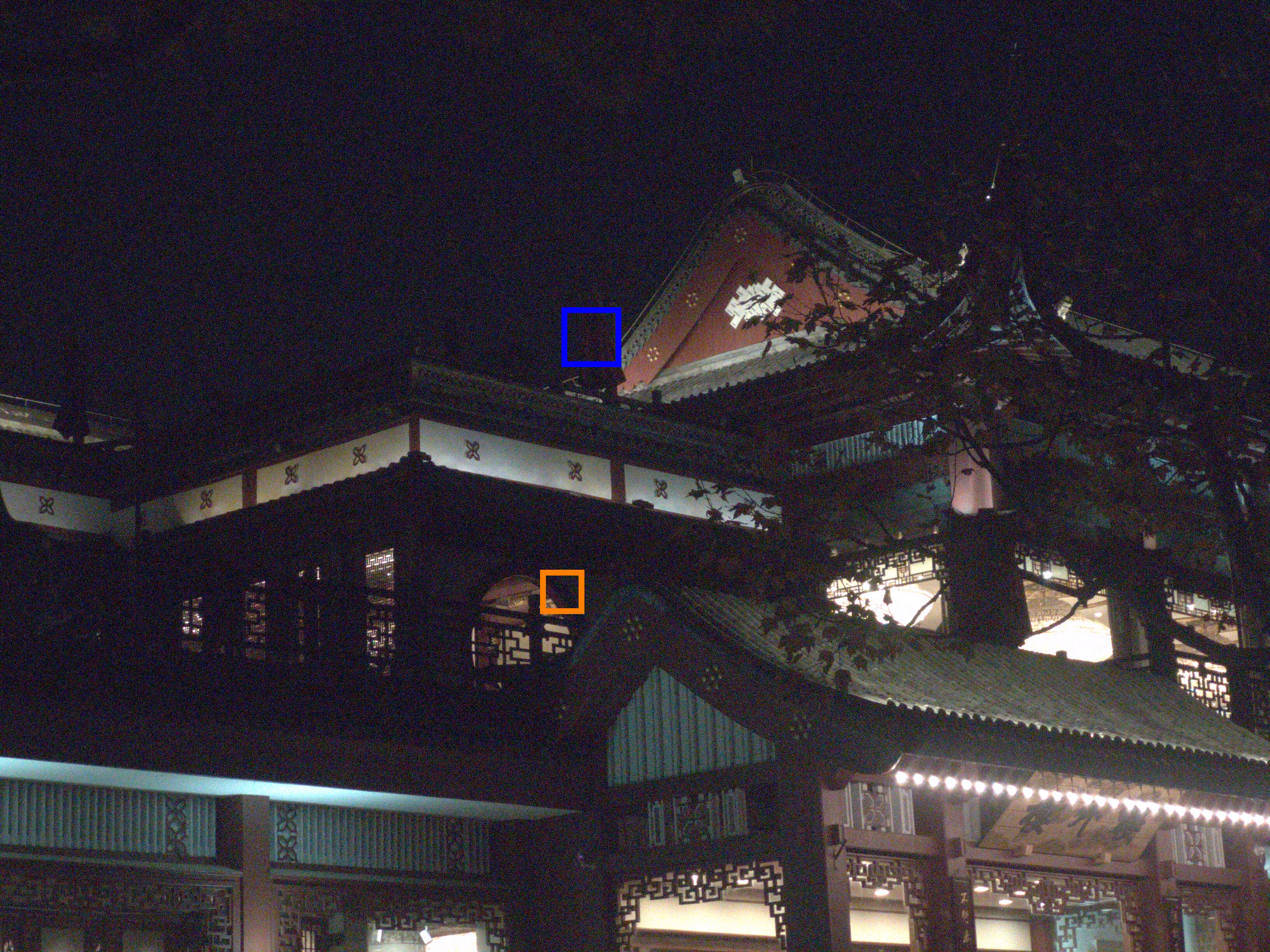}%
    \\[2pt]%
    \includegraphics[width=0.31\linewidth,cframe=blue 1pt]{huawei-0003-noisy-330\_085.png}\hfill%
    \includegraphics[width=0.31\linewidth,cframe=blue 1pt]{huawei-0003-dualdn-330\_085.png}\hfill%
    \includegraphics[width=0.31\linewidth,cframe=blue 1pt]{huawei-0003-25\_15\_9nbr-330\_085.png}%
    \\[2pt]%
    \subcaptionbox{Noisy}{\includegraphics[width=0.31\linewidth,cframe=orange 1pt]{huawei-0003-noisy-275\_760.png}}\hfill%
    \subcaptionbox{DualDn}{\includegraphics[width=0.31\linewidth,cframe=orange 1pt]{huawei-0003-dualdn-275\_760.png}}\hfill%
    \subcaptionbox{Ours}{\includegraphics[width=0.31\linewidth,cframe=orange 1pt]{huawei-0003-25\_15\_9nbr-275\_760.png}}%
    \end{minipage}\hfill%
    \begin{minipage}{0.325\linewidth}
    \subcaptionsetup[figure]{position=top}
    \subcaptionbox{Xiaomi (ISO 3200)}{\includegraphics[width=\linewidth]{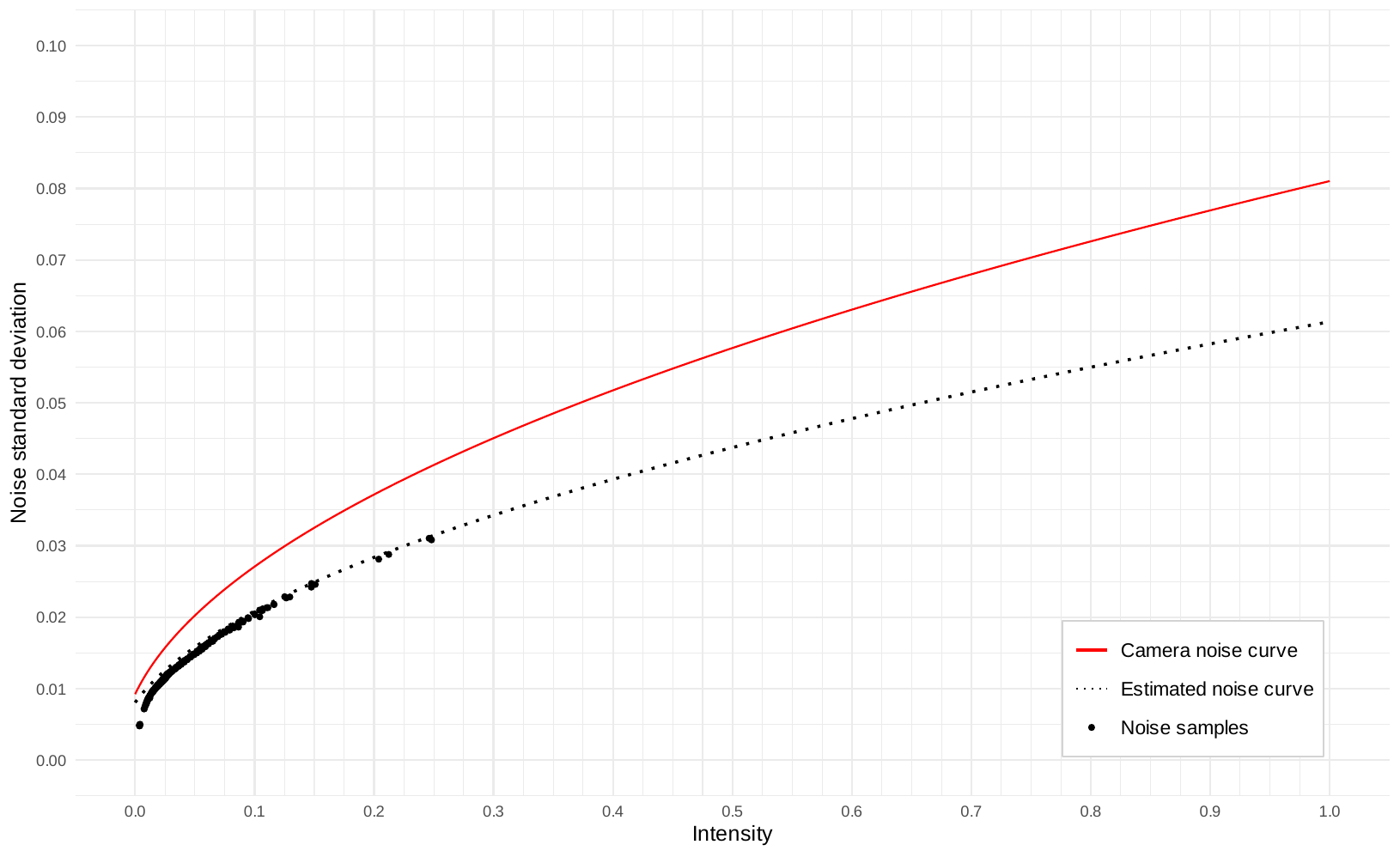}}%
    \\[2pt]%
    \subcaptionsetup[figure]{position=bottom}
    \includegraphics[width=\linewidth]{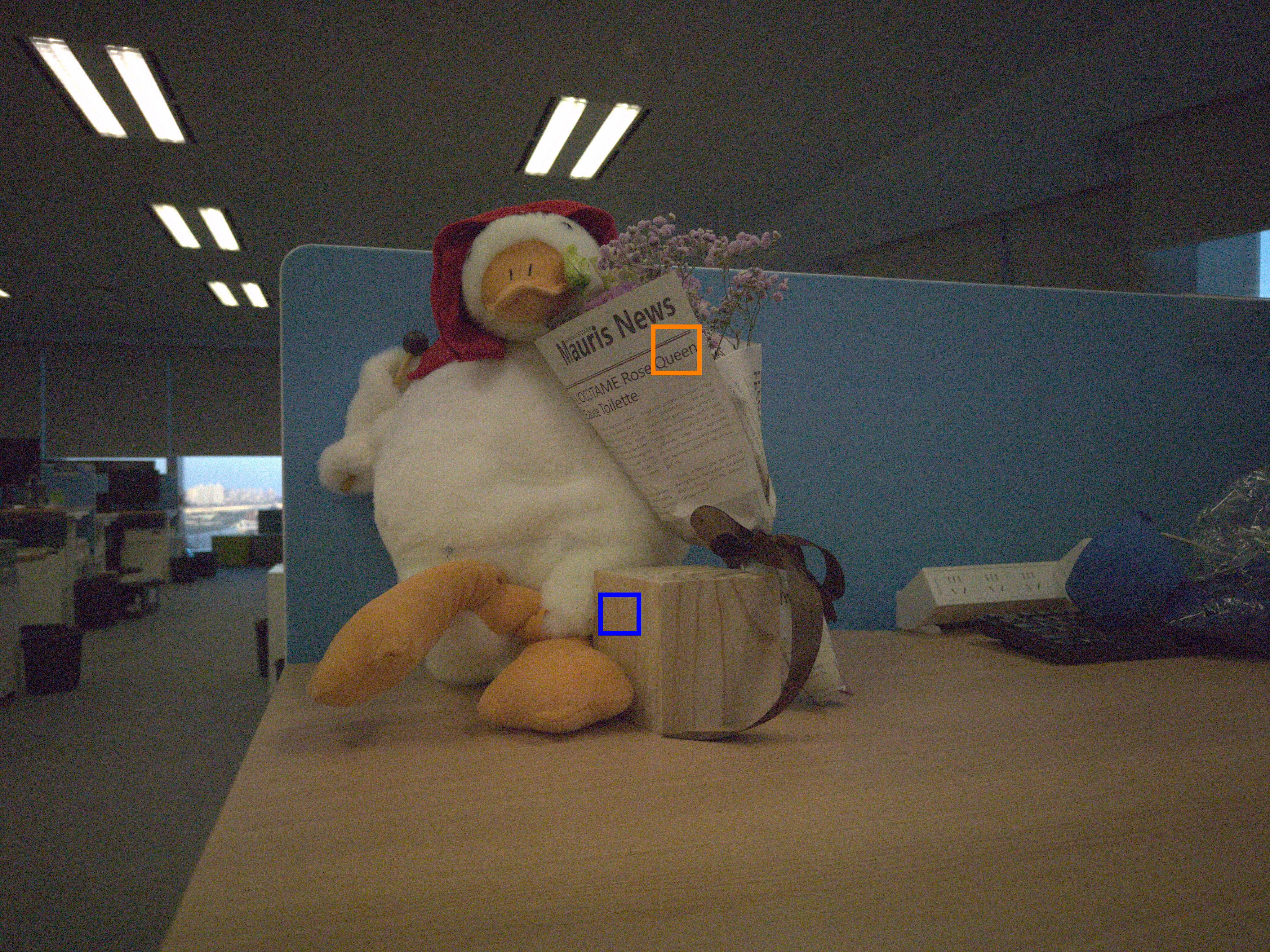}%
    \\[2pt]%
    \includegraphics[width=0.31\linewidth,cframe=blue 1pt]{xiaomi-0001-noisy-400\_895.png}\hfill%
    \includegraphics[width=0.31\linewidth,cframe=blue 1pt]{xiaomi-0001-dualdn-400\_895.png}\hfill%
    \includegraphics[width=0.31\linewidth,cframe=blue 1pt]{xiaomi-0001-25\_15\_9nbr-400\_895.png}%
    \\[2pt]%
    \subcaptionbox{Noisy}{\includegraphics[width=0.31\linewidth,cframe=orange 1pt]{xiaomi-0001-noisy-570\_030.png}}\hfill%
    \subcaptionbox{DualDn}{\includegraphics[width=0.31\linewidth,cframe=orange 1pt]{xiaomi-0001-dualdn-570\_030.png}}\hfill%
    \subcaptionbox{Ours}{\includegraphics[width=0.31\linewidth,cframe=orange 1pt]{xiaomi-0001-25\_15\_9nbr-570\_030.png}}%
    \end{minipage}\hfill%
    \begin{minipage}{0.325\linewidth}
    \subcaptionsetup[figure]{position=top}
    \subcaptionbox{iPhone (ISO 6400)}{\includegraphics[width=\linewidth]{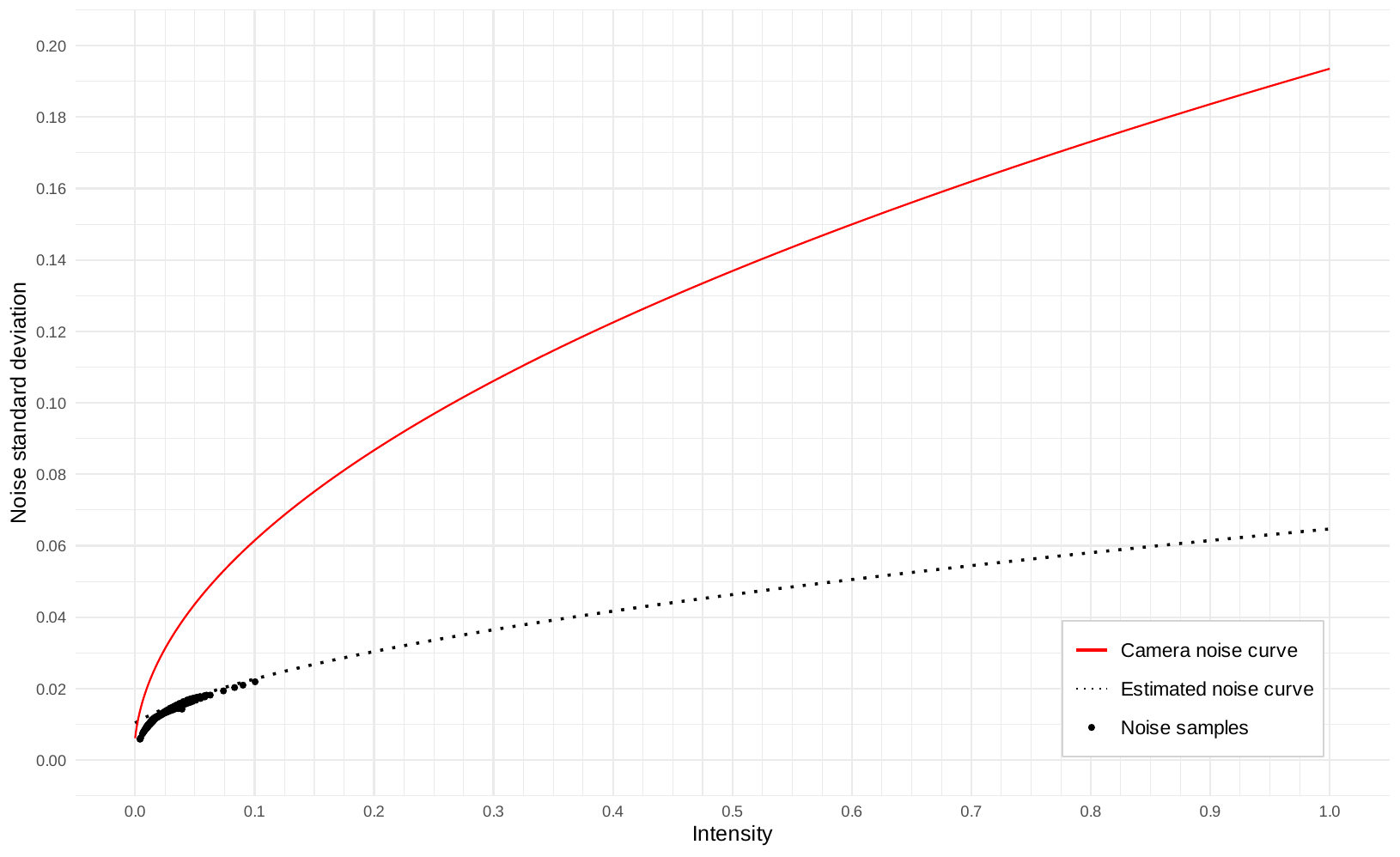}}%
    \\[2pt]%
    \subcaptionsetup[figure]{position=bottom}
    \includegraphics[width=\linewidth]{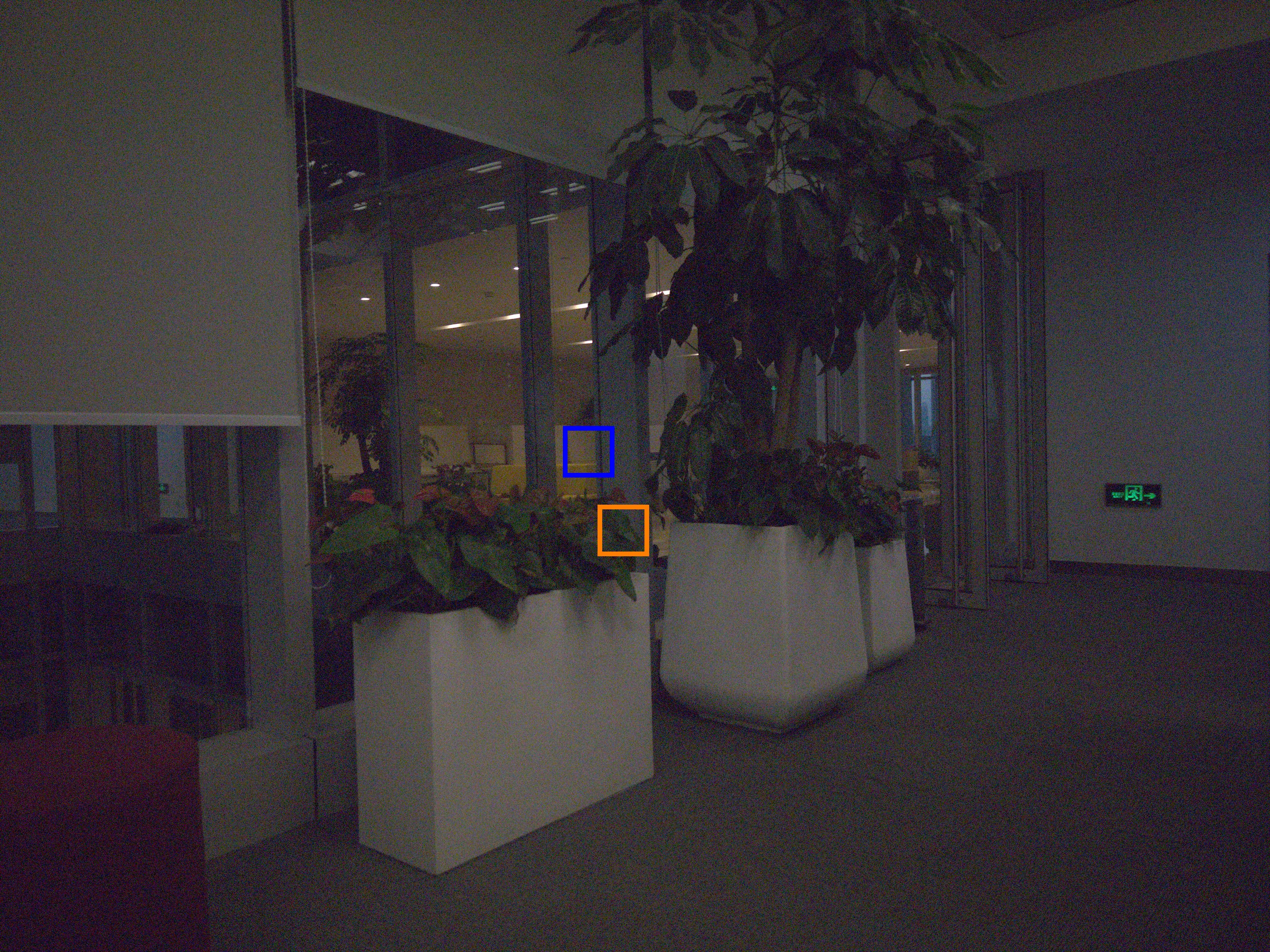}%
    \\[2pt]%
    \includegraphics[width=0.31\linewidth,cframe=blue 1pt]{iphone-0003-noisy-290\_360.png}\hfill%
    \includegraphics[width=0.31\linewidth,cframe=blue 1pt]{iphone-0003-dualdn-290\_360.png}\hfill%
    \includegraphics[width=0.31\linewidth,cframe=blue 1pt]{iphone-0003-25\_15\_9nbr-290\_360.png}%
    \\[2pt]%
    \subcaptionbox{Noisy}{\includegraphics[width=0.31\linewidth,cframe=orange 1pt]{iphone-0003-noisy-400\_610.png}}\hfill%
    \subcaptionbox{DualDn}{\includegraphics[width=0.31\linewidth,cframe=orange 1pt]{iphone-0003-dualdn-400\_610.png}}\hfill%
    \subcaptionbox{Ours}{\includegraphics[width=0.31\linewidth,cframe=orange 1pt]{iphone-0003-25\_15\_9nbr-400\_610.png}}%
    \end{minipage}%
    \caption{Visual comparison on in-the-wild smartphone photographs. Each column shows an image captured with a different device and ISO. First row: noise curve of the image, as embedded in the file metadata (red) and as estimated by us (dotted black). Second row and below: noisy image, and patches denoised with DualDn (RAW) and with our method with 25/15/9 neighbors.}%
    \label{fig:wild}
\end{figure*}

Although the image files contain metadata with noise information at the sensor, we find that the embedded curves are probably miscalibrated and strongly overestimate the level of noise in some sensor settings.
As such, we use the same algorithm as in Section~\ref{sec:adaptationraw} to reestimate the curves for every image.
The first row of Figure~\ref{fig:wild} shows the curves for an image taken with each camera.
The solid red line is constructed from the parameters in the image metadata, while the dotted black one is our estimation from the black points obtained using the method by~\citet{Colom2013Ponom}.
It is apparent that the camera curves are uniformly above our estimations.

\begin{figure*}
    \centering
    \captionsetup[subfigure]{labelformat=empty}
    \begin{minipage}{0.495\linewidth}
    \includegraphics[width=\linewidth]{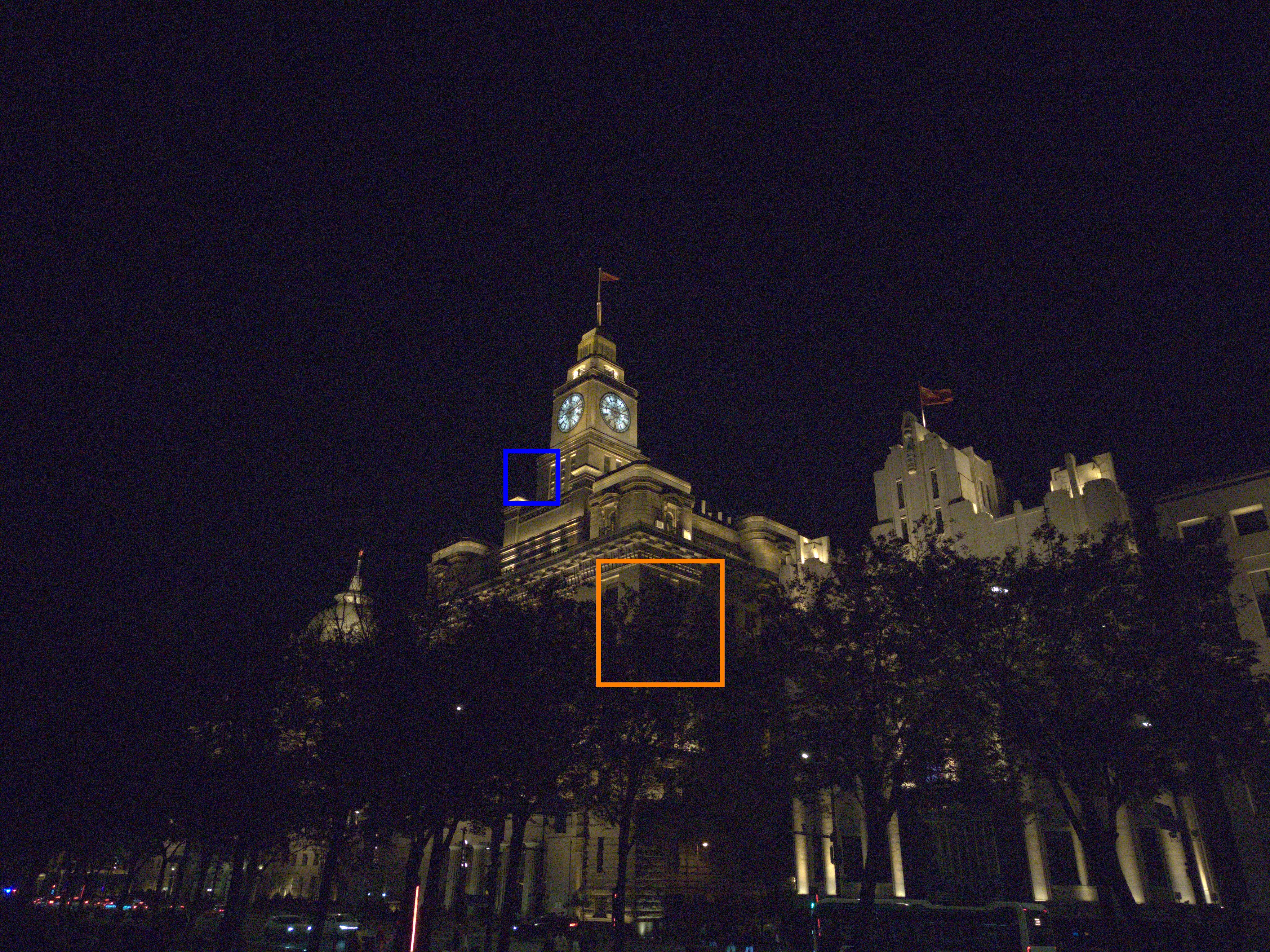}%
    \\[2pt]%
    \includegraphics[width=0.31\linewidth,cframe=blue 1pt]{xiaomi-0006-noisy-1630\_1455.png}\hfill%
    \includegraphics[width=0.31\linewidth,cframe=blue 1pt]{xiaomi-0006-dualdn-1630\_1455.png}\hfill%
    \includegraphics[width=0.31\linewidth,cframe=blue 1pt]{xiaomi-0006-25\_15\_9nbr-1630\_1455.png}%
    \\[2pt]%
    \subcaptionbox{Noisy}{\includegraphics[width=0.31\linewidth,cframe=orange 1pt]{xiaomi-0006-noisy-1930\_1810.png}}\hfill%
    \subcaptionbox{DualDn}{\includegraphics[width=0.31\linewidth,cframe=orange 1pt]{xiaomi-0006-dualdn-1930\_1810.png}}\hfill%
    \subcaptionbox{Ours}{\includegraphics[width=0.31\linewidth,cframe=orange 1pt]{xiaomi-0006-25\_15\_9nbr-1930\_1810.png}}%
    \end{minipage}\hfill%
    \begin{minipage}{0.495\linewidth}
    \includegraphics[width=\linewidth]{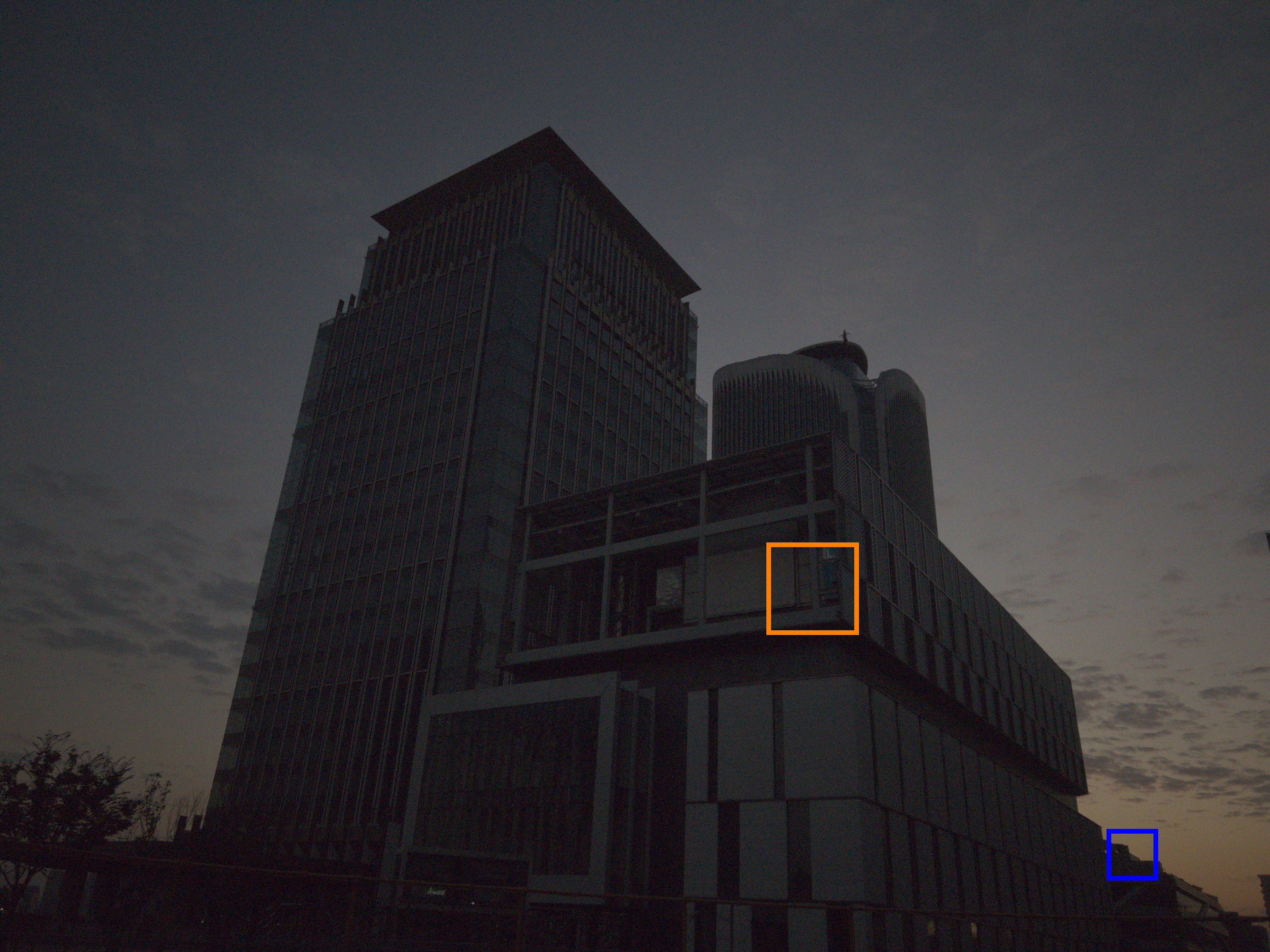}%
    \\[2pt]%
    \includegraphics[width=0.31\linewidth,cframe=blue 1pt]{iphone-0004-noisy-3520\_2640.png}\hfill%
    \includegraphics[width=0.31\linewidth,cframe=blue 1pt]{iphone-0004-dualdn-3520\_2640.png}\hfill%
    \includegraphics[width=0.31\linewidth,cframe=blue 1pt]{iphone-0004-25\_15\_9nbr-3520\_2640.png}%
    \\[2pt]%
    \subcaptionbox{Noisy}{\includegraphics[width=0.31\linewidth,cframe=orange 1pt]{iphone-0004-noisy-2440\_1730.png}}\hfill%
    \subcaptionbox{DualDn}{\includegraphics[width=0.31\linewidth,cframe=orange 1pt]{iphone-0004-dualdn-2440\_1730.png}}\hfill%
    \subcaptionbox{Ours}{\includegraphics[width=0.31\linewidth,cframe=orange 1pt]{iphone-0004-25\_15\_9nbr-2440\_1730.png}}%
    \end{minipage}%
    \caption{Visual comparison between DualDn (RAW) and our method with 25/15/9 neighbors on low-light outdoor photographs. DualDn leaves residual grain and low-frequency noise on dark flat regions, and shows a slight color cast around dark textures, which we successfully correct.}%
    \label{fig:wild2}
\end{figure*}

Below the plots in Figure~\ref{fig:wild}, we showcase the results of denoising the images with DualDn (26.5M parameters) and our method (15.3M parameters) using the estimated black dotted curves.
As the commercial ISP on the devices is unknown, we apply a simple processing pipeline using the EXIF metadata for the purpose of visualization.
The Huawei scene is notably dark and thus serves to highlight the behavior of both methods in low-light conditions.
DualDn exhibits difficulties in such dark regions: blotchy artifacts appear around edges, blue-tinted low-frequency residual noise remains in flatter areas, and blue hot-pixel defects are visible.
In the blue patch, the parasol loses its silhouette and blends into the dark background; in the orange patch, a blotch appears along the edge of the archway.
Our method produces a cleaner reconstruction in both cases.
On the Xiaomi image, we achieve finer texture preservation overall: DualDn oversmooths the wood markings and introduces yellowish artifacts around the newspaper text, whereas we recover sharper forms.
Finally, on the iPhone photograph, DualDn leaves noticeable residual noise on flat regions.
While this underfiltering occasionally preserves local contrast (e.g.\ the edges of the plant appear slightly sharper), the persistent noise degrades the overall visual quality compared to our cleaner restoration.

Figure~\ref{fig:wild2} displays more examples of very low-light photographs from the same dataset, taken in outdoor settings with high ISO.
Our method manages to successfully remove the noise everywhere with minimal texture loss, yielding visually faithful results even in regions with very low signal-to-noise ratio.
This is noticeable in the orange patch of the left image, where noise obfuscates most of the trees.
Where DualDn enshrouds them with a slight color bias in the near-black, we get a sharper reconstruction of the branches with a more cohesive color.
Differences between how both methods treat dark regions can also be seen in the blue patch of the same image: DualDn leaves some mix of grain and low-frequency noise in the sky and in the wall of the building, which we completely filter.
This residual noise left by DualDn is mostly found in dark, flat regions, as further showcased in both patches of the right image, and follows the patterns already seen in the third image of Figure~\ref{fig:wild} in an indoor setting.
Our method again overcomes this issue, fully removing the noise in such regions without compromising texture fidelity.

Overall, these experiments confirm that the proposed method generalizes well across sensors, lighting conditions and ISO levels, yielding visually consistent, sharp results without residual noise or excessive texture smoothing, and being competitive with larger, more complex (and less interpretable) RAW denoising pipelines in terms of image quality---traits that make it suitable for mobile photography.

\section{Conclusion}

We have presented a neural network architecture that translates the classical three-step pipeline of nonlocal patch-based image denoising methods (matching, collaborative filtering, and aggregation) into a fully learnable framework operating on feature representations.
The centerpiece of the network is an interpretable Nonlocal Feature Matching and Filtering Block, which uses a convolutional neural network to select a fixed number of neighbors as support for subsequent filtering through the modulation of a groupwise lineally transformed feature stack.
This block is placed within a UNet, efficiently expanding the receptive field in a nonlocal manner without requiring expensive self-attention or excessive depth.

By means of training data selection and using a noise level map as input, the proposed network is designed for the task of RAW-to-RAW image denoising.
We have curated a dataset of clean RAW images alongside a list of noise profiles for a variety of camera sensors and ranges of ISO values, which we use to synthesize realistic RAW images corrupted with Poisson-Gaussian noise.
Training the network on this data yields a sensor-agnostic denoiser that generalizes well to unseen devices and lighting conditions, as evidenced by visual results on in-the-wild photographs.
Quantitative experiments also show that the proposed method achieves results competitive with state-of-the-art CNN and transformer-based denoisers while using significantly fewer parameters.

\section*{Acknowledgments}
The authors gratefully acknowledge the computer resources at Artemisa and the technical support provided by the Instituto de Fisica Corpuscular, IFIC (CSIC-UV). Artemisa is co-funded by the European Union through the 2014-2020 ERDF Operative Programme of Comunitat Valenciana, project IDIFEDER/2018/048.

This work was funded by MCIN/AEI/10.13039/501100011033 and by \enquote{ERDF A way of making Europe}, European Union, under grant PID2021-1257110B-I00.
The work of Marco Sánchez-Beeckman was also supported by the Conselleria de Fons Europeus, Universitat i Cultura del Govern de les Illes Balears under grant FPU2023-011-C.

\printcredits

\bibliographystyle{cas-model2-names} 
\bibliography{manuscript}

@article{Buades2020CFA,
    author = {Buades, Antoni and Duran, Joan},
    journal = {IEEE Transactions on Circuits and Systems for Video Technology},
    title = {{CFA} Video Denoising and Demosaicking Chain via Spatio-Temporal
             Patch-Based Filtering},
    year = {2020},
    volume = {30},
    number = {11},
    pages = {4143-4157},
    doi = {10.1109/TCSVT.2019.2956691},
}

@article{Sanchez2025Combining,
  author={Sánchez-Beeckman, M. and Buades, A. and Brandonisio, N. and Kanoun, B.},
  journal={IEEE Transactions on Image Processing}, 
  title={Combining Pre- and Post-Demosaicking Noise Removal for {RAW} Video}, 
  year={2026},
  volume={35},
  number={},
  pages={1652-1667},
  doi={10.1109/TIP.2025.3527886}
}

@inproceedings{Akiyama2015Pseudo,
    author={Akiyama, Hiroki and Tanaka, Masayuki and Okutomi, Masatoshi},
    booktitle={IEEE International Conference on Image Processing (ICIP)}, 
    title={Pseudo four-channel image denoising for noisy {CFA} raw data}, 
    year={2015},
    volume={},
    number={},
    pages={4778-4782},
    doi={10.1109/ICIP.2015.7351714}
}

@article{Facciolo2017Multiscale,
    author = {Facciolo, Gabriele and Pierazzo, Nicola and Morel, Jean-Michel},
    title = {Conservative Scale Recomposition for Multiscale Denoising (The Devil is in the High Frequency Detail)},
    journal = {SIAM Journal on Imaging Sciences},
    volume = {10},
    number = {3},
    pages = {1603-1626},
    year = {2017},
    doi = {10.1137/17M1111826},
}

@inproceedings{Burger2011Multiscale,
    author={Burger, Harold Christopher and Harmeling, Stefan},
    editor={Mester, Rudolf and Felsberg, Michael},
    title={Improving Denoising Algorithms via a Multi-scale Meta-procedure},
    booktitle={Joint Pattern Recognition Symposium},
    year={2011},
    publisher={Springer Berlin Heidelberg},
    pages={206--215},
    doi={10.1007/978-3-642-23123-0_21}
}

@inproceedings{Ronneberger2015UNet,
    author={Ronneberger, Olaf and Fischer, Philipp and Brox, Thomas},
    editor={Navab, Nassir and Hornegger, Joachim and Wells, William M. and Frangi, Alejandro F.},
    title={{U}-{N}et: Convolutional Networks for Biomedical Image Segmentation},
    booktitle={Medical Image Computing and Computer-Assisted Intervention (MICCAI)},
    year={2015},
    publisher={Springer International Publishing},
    address={Cham},
    pages={234--241},
    doi={10.1007/978-3-319-24574-4_28}
}

@inproceedings{He2016ResNet,
    author={He, Kaiming and Zhang, Xiangyu and Ren, Shaoqing and Sun, Jian},
    booktitle={IEEE Conference on Computer Vision and Pattern Recognition (CVPR)}, 
    title={Deep Residual Learning for Image Recognition}, 
    year={2016},
    volume={},
    number={},
    pages={770-778},
    doi={10.1109/CVPR.2016.90}
}

@inproceedings{Buades2005NLMeans,
    author={Buades, Antoni and Coll, Bartomeu and Morel, Jean-Michel},
    booktitle={IEEE Computer Society Conference on Computer Vision and Pattern Recognition (CVPR)}, 
    title={A non-local algorithm for image denoising}, 
    year={2005},
    volume={2},
    number={},
    pages={60-65},
    doi={10.1109/CVPR.2005.38}
}

@article{Dabov2007BM3D,
    author={Dabov, Kostadin and Foi, Alessandro and Katkovnik, Vladimir and Egiazarian, Karen},
    journal={IEEE Transactions on Image Processing}, 
    title={Image Denoising by Sparse {3-D} Transform-Domain Collaborative Filtering}, 
    year={2007},
    volume={16},
    number={8},
    pages={2080-2095},
    publisher = {IEEE},
    doi={10.1109/TIP.2007.901238}
}

@article{Lebrun2013NLBayes,
    title = {A nonlocal {B}ayesian image denoising algorithm},
    author = {Lebrun, Marc and Buades, Antoni and Morel, Jean-Michel},
    journal = {SIAM Journal on Imaging Sciences},
    volume = {6},
    number = {3},
    pages = {1665--1688},
    year = {2013},
    publisher = {SIAM},
    doi = {10.1137/120874989}
}

@article{Lebrun2015Multiscale,
    author={Lebrun, Marc and Colom, Miguel and Morel, Jean-Michel},
    journal={IEEE Transactions on Image Processing}, 
    title={Multiscale Image Blind Denoising}, 
    year={2015},
    volume={24},
    number={10},
    pages={3149-3161},
    doi={10.1109/TIP.2015.2439041}
}

@inproceedings{Liu2022ConvNeXt,
    author={Liu, Zhuang and Mao, Hanzi and Wu, Chao-Yuan and Feichtenhofer, Christoph and Darrell, Trevor and Xie, Saining},
    booktitle={IEEE/CVF Conference on Computer Vision and Pattern Recognition (CVPR)}, 
    title={A ConvNet for the 2020s}, 
    year={2022},
    volume={},
    number={},
    pages={11966-11976},
    doi={10.1109/CVPR52688.2022.01167}
}

@article{Kervrann2006Optimal,
    author={Kervrann, Charles and Boulanger, J{\'e}r{\^o}me},
    journal={IEEE Transactions on Image Processing}, 
    title={Optimal Spatial Adaptation for Patch-Based Image Denoising}, 
    year={2006},
    volume={15},
    number={10},
    pages={2866-2878},
    publisher = {IEEE},
    doi={10.1109/TIP.2006.877529}
}

@article{Elad2023Survey,
    author = {Elad, Michael and Kawar, Bahjat and Vaksman, Gregory},
    title = {Image Denoising: The Deep Learning Revolution and Beyond—A Survey Paper},
    journal = {SIAM Journal on Imaging Sciences},
    volume = {16},
    number = {3},
    pages = {1594-1654},
    year = {2023},
    doi = {10.1137/23M1545859},
}

@article{Milanfar2013Tour,
    author={Milanfar, Peyman},
    journal={IEEE Signal Processing Magazine}, 
    title={A Tour of Modern Image Filtering: New Insights and Methods, Both Practical and Theoretical}, 
    year={2013},
    volume={30},
    number={1},
    pages={106-128},
    doi={10.1109/MSP.2011.2179329}
}

@inproceedings{Agustsson2017DIV2K,
    author={Agustsson, Eirikur and Timofte, Radu},
    booktitle={IEEE Conference on Computer Vision and Pattern Recognition Workshops (CVPRW)}, 
    title={{NTIRE} 2017 Challenge on Single Image Super-Resolution: Dataset and Study}, 
    year={2017},
    volume={},
    number={},
    pages={1122-1131},
    doi={10.1109/CVPRW.2017.150}
}

@article{Ma2017WED,
    author={Ma, Kede and Duanmu, Zhengfang and Wu, Qingbo and Wang, Zhou and Yong, Hongwei and Li, Hongliang and Zhang, Lei},
    journal={IEEE Transactions on Image Processing}, 
    title={Waterloo Exploration Database: New Challenges for Image Quality Assessment Models}, 
    year={2017},
    volume={26},
    number={2},
    pages={1004-1016},
    doi={10.1109/TIP.2016.2631888}
}

@inproceedings{Lim2017Flickr2K,
    author={Lim, Bee and Son, Sanghyun and Kim, Heewon and Nah, Seungjun and Lee, Kyoung Mu},
    booktitle={IEEE Conference on Computer Vision and Pattern Recognition Workshops (CVPRW)}, 
    title={Enhanced Deep Residual Networks for Single Image Super-Resolution}, 
    year={2017},
    volume={},
    number={},
    pages={1132-1140},
    doi={10.1109/CVPRW.2017.151}
}

@inproceedings{Chen2021IPT,
    author={Chen, Hanting and Wang, Yunhe and Guo, Tianyu and Xu, Chang and Deng, Yiping and Liu, Zhenhua and Ma, Siwei and Xu, Chunjing and Xu, Chao and Gao, Wen},
    booktitle={IEEE/CVF Conference on Computer Vision and Pattern Recognition (CVPR)}, 
    title={Pre-Trained Image Processing Transformer}, 
    year={2021},
    volume={},
    number={},
    pages={12294-12305},
    doi={10.1109/CVPR46437.2021.01212}
}

@article{Yin2022CSformer,
    author={Yin, Haitao and Ma, Siyuan},
    journal={IEEE Signal Processing Letters}, 
    title={{CSformer}: Cross-Scale Features Fusion Based Transformer for Image Denoising}, 
    year={2022},
    volume={29},
    number={},
    pages={1809-1813},
    doi={10.1109/LSP.2022.3199145}
}

@article{Li2024EWT,
    title = {{EWT}: Efficient Wavelet-Transformer for single image denoising},
    journal = {Neural Networks},
    volume = {177},
    pages = {106378},
    year = {2024},
    issn = {0893-6080},
    doi = {10.1016/j.neunet.2024.106378},
    author = {Juncheng Li and Bodong Cheng and Ying Chen and Guangwei Gao and Jun Shi and Tieyong Zeng},
}

@inproceedings{Zhou2024Efficient,
    author = {Zhou, Yubo and Lin, Jin and Ye, Fangchen and Qu, Yanyun and Xie, Yuan},
    title = {Efficient lightweight image denoising with triple attention transformer},
    year = {2024},
    isbn = {978-1-57735-887-9},
    publisher = {AAAI Press},
    doi = {10.1609/aaai.v38i7.28604},
    booktitle = {AAAI Conference on Artificial Intelligence},
    articleno = {856},
    numpages = {9},
    pages={7704-7712}
}

@article{Zhuge2023FeatureEnhanced,
    title = {Single image denoising with a feature-enhanced network},
    journal = {Neural Networks},
    volume = {168},
    pages = {313-325},
    year = {2023},
    issn = {0893-6080},
    doi = {10.1016/j.neunet.2023.08.056},
    author = {Ruibin Zhuge and Jinghua Wang and Zenglin Xu and Yong Xu},
}

@inproceedings{Liu2021Swin,
  author={Liu, Ze and Lin, Yutong and Cao, Yue and Hu, Han and Wei, Yixuan and Zhang, Zheng and Lin, Stephen and Guo, Baining},
  booktitle={IEEE/CVF International Conference on Computer Vision (ICCV)}, 
  title={Swin Transformer: Hierarchical Vision Transformer using Shifted Windows}, 
  year={2021},
  volume={},
  number={},
  pages={9992-10002},
  doi={10.1109/ICCV48922.2021.00986}
}

@inproceedings{Liang2021SwinIR,
    author={Liang, Jingyun and Cao, Jiezhang and Sun, Guolei and Zhang, Kai and Van Gool, Luc and Timofte, Radu},
    booktitle={IEEE/CVF International Conference on Computer Vision Workshops (ICCVW)}, 
    title={{SwinIR}: Image Restoration Using Swin Transformer}, 
    year={2021},
    volume={},
    number={},
    pages={1833-1844},
    doi={10.1109/ICCVW54120.2021.00210}
}

@inproceedings{Wang2022Uformer,
    author={Wang, Zhendong and Cun, Xiaodong and Bao, Jianmin and Zhou, Wengang and Liu, Jianzhuang and Li, Houqiang},
    booktitle={IEEE/CVF Conference on Computer Vision and Pattern Recognition (CVPR)}, 
    title={Uformer: A General {U}-Shaped Transformer for Image Restoration}, 
    year={2022},
    volume={},
    number={},
    pages={17662-17672},
    doi={10.1109/CVPR52688.2022.01716}
}

@inproceedings{Zhang2024Xformer,
    title={Xformer: Hybrid {X}-Shaped Transformer for Image Denoising}, 
    author={Jiale Zhang and Yulun Zhang and Jinjin Gu and Jiahua Dong and Linghe Kong and Xiaokang Yang},
    booktitle={International Conference on Learning Representations},
    year={2024},
}

@article{Hu2026DSCA,
    author = {Yuxuan Hu and Debo Cheng and Zhirong Huang and Boyan Chen and Shilong Lin and Shichao Zhang},
    title = {{DSCA}-former: Dual-stem cross-attentive transformer for image denoising},
    journal = {Neurocomputing},
    pages = {132656},
    year = {2026},
    issn = {0925-2312},
    doi = {10.1016/j.neucom.2026.132656},
}

@inproceedings{NTIRE2025,
  author={Sun, Lei and Guo, Hang and Ren, Bin and Van Gool, Luc and Timofte, Radu and Li, Yawei},
  booktitle={IEEE/CVF Conference on Computer Vision and Pattern Recognition Workshops (CVPRW)}, 
  title={The Tenth {NTIRE} 2025 Image Denoising Challenge Report}, 
  year={2025},
  volume={},
  number={},
  pages={1333-1360},
  doi={10.1109/CVPRW67362.2025.00125}
}

@article{Zhang2022DPIR,
    author={Zhang, Kai and Li, Yawei and Zuo, Wangmeng and Zhang, Lei and Van Gool, Luc and Timofte, Radu},
    journal={IEEE Transactions on Pattern Analysis and Machine Intelligence}, 
    title={Plug-and-Play Image Restoration With Deep Denoiser Prior}, 
    year={2022},
    volume={44},
    number={10},
    pages={6360-6376},
    doi={10.1109/TPAMI.2021.3088914}
}

@inproceedings{Liang2022RVRT,
    author = {Liang, Jingyun and Fan, Yuchen and Xiang, Xiaoyu and Ranjan, Rakesh and Ilg, Eddy and Green, Simon and Cao, Jiezhang and Zhang, Kai and Timofte, Radu and Van Gool, Luc},
    title = {Recurrent video restoration transformer with guided deformable attention},
    year = {2022},
    isbn = {9781713871088},
    publisher = {Curran Associates Inc.},
    address = {Red Hook, NY, USA},
    booktitle = {International Conference on Neural Information Processing Systems (NeurIPS)},
    editor = {S. Koyejo and S. Mohamed and A. Agarwal and D. Belgrave and K. Cho and A. Oh},
    articleno = {28},
    numpages = {16},
    volume = {35},
    pages = {378--393},
    location = {New Orleans, LA, USA},
    doi = {10.5555/3600270.3600298}
}

@article{Zhang2018FFDNet,
    author={Zhang, Kai and Zuo, Wangmeng and Zhang, Lei},
    journal={IEEE Transactions on Image Processing}, 
    title={{FFDNet}: Toward a Fast and Flexible Solution for {CNN}-Based Image Denoising}, 
    year={2018},
    volume={27},
    number={9},
    pages={4608-4622},
    doi={10.1109/TIP.2018.2839891}
}

@article{Yan2020Combining,
  author={Yan, Zifei and Guo, Shi and Xiao, Gang and Zhang, Hongzhi},
  journal={IEEE Access}, 
  title={On Combining {CNN} With Non-Local Self-Similarity Based Image Denoising Methods}, 
  year={2020},
  volume={8},
  number={},
  pages={14789-14797},
  doi={10.1109/ACCESS.2019.2962809}
}

@inproceedings{Zamir2020CycleISP,
    author={Zamir, Syed Waqas and Arora, Aditya and Khan, Salman and Hayat, Munawar and Khan, Fahad Shahbaz and Yang, Ming-Hsuan and Shao, Ling},
    booktitle={IEEE/CVF Conference on Computer Vision and Pattern Recognition (CVPR)}, 
    title={{CycleISP}: Real Image Restoration via Improved Data Synthesis}, 
    year={2020},
    volume={},
    number={},
    pages={2693-2702},
    doi={10.1109/CVPR42600.2020.00277}
}

@article{Cao2024PseudoISP,
    author = {Yue Cao and Xiaohe Wu and Shuran Qi and Xiao Liu and Zhongqin Wu and Wangmeng Zuo},
    title = {Pseudo-{ISP}: Learning pseudo in-camera signal processing pipeline from a color image denoiser},
    journal = {Neurocomputing},
    volume = {605},
    pages = {128316},
    year = {2024},
    issn = {0925-2312},
    doi = {10.1016/j.neucom.2024.128316}
}

@inproceedings{Huang2015Urban100,
    author={Huang, Jia-Bin and Singh, Abhishek and Ahuja, Narendra},
    booktitle={IEEE Conference on Computer Vision and Pattern Recognition (CVPR)}, 
    title={Single image super-resolution from transformed self-exemplars}, 
    year={2015},
    volume={},
    number={},
    pages={5197-5206},
    doi={10.1109/CVPR.2015.7299156}
}

@misc{Franzen1999Kodak,
    title={Kodak lossless true color image suite},
    author={Franzen, Rich},
    year={1999},
    url={https://r0k.us/graphics/kodak/}, 
}

@article{Zhang2011McMaster,
    title={Color demosaicking by local directional interpolation and nonlocal adaptive thresholding},
    author={Zhang, Lei and Wu, Xiaolin and Buades, Antoni and Li, Xin},
    journal={Journal of Electronic imaging},
    volume={20},
    number={2},
    pages={023016--023016},
    year={2011},
    publisher={Society of Photo-Optical Instrumentation Engineers},
    doi={10.1117/1.3600632},
}

@inproceedings{Martin2001CBSD68,
    author={Martin, D. and Fowlkes, C. and Tal, D. and Malik, J.},
    booktitle={IEEE International Conference on Computer Vision (ICCV)}, 
    title={A database of human segmented natural images and its application to evaluating segmentation algorithms and measuring ecological statistics}, 
    year={2001},
    volume={2},
    number={},
    pages={416-423},
    doi={10.1109/ICCV.2001.937655}
}

@inproceedings{Plotz2017DND,
    author={Pl\"{o}tz, Tobias and Roth, Stefan},
    booktitle={IEEE Conference on Computer Vision and Pattern Recognition (CVPR)}, 
    title={Benchmarking Denoising Algorithms with Real Photographs}, 
    year={2017},
    volume={},
    number={},
    pages={2750-2759},
    doi={10.1109/CVPR.2017.294}
}

@misc{Brummer2025RawNIND,
    title={Learning Joint Denoising, Demosaicing, and Compression from the Raw Natural Image Noise Dataset}, 
    author={Benoit Brummer and Christophe De Vleeschouwer},
    year={2025},
    eprint={2501.08924},
    archivePrefix={arXiv},
    primaryClass={cs.CV},
    url={https://arxiv.org/abs/2501.08924}, 
}

@inproceedings{Li2024DualDn,
    author = {Li, Ruikang and Wang, Yujin and Chen, Shiqi and Zhang, Fan and Gu, Jinwei and Xue, Tianfan},
    editor={Leonardis, Ale{\v{s}} and Ricci, Elisa and Roth, Stefan and Russakovsky, Olga and Sattler, Torsten and Varol, G{\"u}l},
    title = {{DualDn}: Dual-Domain Denoising via Differentiable {ISP}},
    year = {2024},
    isbn = {978-3-031-73635-3},
    publisher = {Springer-Verlag},
    address = {Berlin, Heidelberg},
    doi = {10.1007/978-3-031-73636-0_10},
    booktitle = {European Conference on Computer Vision (ECCV)},
    pages = {160–177},
    numpages = {18},
    location = {Milan, Italy}
}

@inproceedings{Wang2020Practical,
    author = {Wang, Yuzhi and Huang, Haibin and Xu, Qin and Liu, Jiaming and Liu, Yiqun and Wang, Jue},
    editor={Vedaldi, Andrea and Bischof, Horst and Brox, Thomas and Frahm, Jan-Michael},
    title = {Practical Deep Raw Image Denoising on Mobile Devices},
    year = {2020},
    isbn = {978-3-030-58538-9},
    publisher = {Springer-Verlag},
    address = {Berlin, Heidelberg},
    doi = {10.1007/978-3-030-58539-6_1},
    booktitle={European Conference on Computer Vision (ECCV)},
    pages = {1-–16},
    numpages = {16},
    location = {Glasgow, United Kingdom}
}

@inproceedings{Chang2020SADNet,
    author={Chang, Meng and Li, Qi and Feng, Huajun and Xu, Zhihai},
    editor={Vedaldi, Andrea and Bischof, Horst and Brox, Thomas and Frahm, Jan-Michael},
    title={Spatial-Adaptive Network for Single Image Denoising},
    booktitle={European Conference on Computer Vision (ECCV)},
    year={2020},
    publisher={Springer International Publishing},
    address={Cham},
    pages={171--187},
    doi={10.1007/978-3-030-58577-8_11}
}

@inproceedings{Gou2020CLEARER,
    author = {Gou, Yuanbiao and Li, Boyun and Liu, Zitao and Yang, Songfan and Peng, Xi},
    title = {{CLEARER}: multi-scale neural architecture search for image restoration},
    year = {2020},
    isbn = {9781713829546},
    publisher = {Curran Associates Inc.},
    address = {Red Hook, NY, USA},
    booktitle = {International Conference on Neural Information Processing Systems (NeurIPS)},
    editor = {H. Larochelle and M. Ranzato and R. Hadsell and M.F. Balcan and H. Lin},
    articleno = {1437},
    numpages = {12},
    volume = {33},
    pages = {17129--17140},
    location = {Vancouver, BC, Canada},
    doi = {10.5555/3495724.3497161}
}

@inproceedings{Ren2021DeamNet,
    author={Ren, Chao and He, Xiaohai and Wang, Chuncheng and Zhao, Zhibo},
    booktitle={IEEE/CVF Conference on Computer Vision and Pattern Recognition (CVPR)}, 
    title={Adaptive Consistency Prior based Deep Network for Image Denoising}, 
    year={2021},
    volume={},
    number={},
    pages={8592-8602},
    doi={10.1109/CVPR46437.2021.00849}
}

@inproceedings{Liu2018NLRN,
    author = {Liu, Ding and Wen, Bihan and Fan, Yuchen and Loy, Chen Change and Huang, Thomas S.},
    title = {Non-local recurrent network for image restoration},
    year = {2018},
    volume = {31},
    publisher = {Curran Associates Inc.},
    editor = {S. Bengio and H. Wallach and H. Larochelle and K. Grauman and N. Cesa-Bianchi and R. Garnett},
    address = {Red Hook, NY, USA},
    booktitle = {International Conference on Neural Information Processing Systems (NeurIPS)},
    pages = {1680–1689},
    numpages = {10},
    location = {Montr\'{e}al, Canada},
    doi = {10.5555/3326943.3327097}
}

@inproceedings{Wang2018NLNeural,
    author={Wang, Xiaolong and Girshick, Ross and Gupta, Abhinav and He, Kaiming},
    booktitle={IEEE/CVF Conference on Computer Vision and Pattern Recognition (CVPR)}, 
    title={Non-local Neural Networks}, 
    year={2018},
    volume={},
    number={},
    pages={7794-7803},
    doi={10.1109/CVPR.2018.00813}
}

@inproceedings{Zhang2019RNAN,
    title={Residual Non-local Attention Networks for Image Restoration},
    author={Yulun Zhang and Kunpeng Li and Kai Li and Bineng Zhong and Yun Fu},
    booktitle={International Conference on Learning Representations},
    year={2019},
}

@inproceedings{Plotz2018N3Net,
    author = {Pl\"{o}tz, Tobias and Roth, Stefan},
    title = {Neural nearest neighbors networks},
    year = {2018},
    volume = {31},
    publisher = {Curran Associates Inc.},
    editor = {S. Bengio and H. Wallach and H. Larochelle and K. Grauman and N. Cesa-Bianchi and R. Garnett},
    address = {Red Hook, NY, USA},
    booktitle = {International Conference on Neural Information Processing Systems (NeurIPS)},
    pages = {1095–1106},
    numpages = {12},
    location = {Montr\'{e}al, Canada},
    doi = {10.5555/3326943.3327044}
}

@article{Mei2023PANet,
    title = {Pyramid Attention Network for Image Restoration},
    author = {Yiqun Mei and Yuchen Fan and Yulun Zhang and Jiahui Yu and Yuqian Zhou and Ding Liu and Yun Fu and Thomas S. Huang and Humphrey Shi},
    journal = {International Journal of Computer Vision},
    volume = {131},
    pages = {3207--3225},
    year = {2023},
    doi = {10.1007/s11263-023-01843-5}
}

@inproceedings{Zamir2022Restormer,
    author={Zamir, Syed Waqas and Arora, Aditya and Khan, Salman and Hayat, Munawar and Khan, Fahad Shahbaz and Yang, Ming–Hsuan},
    booktitle={IEEE/CVF Conference on Computer Vision and Pattern Recognition (CVPR)}, 
    title={Restormer: Efficient Transformer for High-Resolution Image Restoration}, 
    year={2022},
    volume={},
    number={},
    pages={5718-5729},
    doi={10.1109/CVPR52688.2022.00564}
}

@inproceedings{Gou2022MSANet,
    author = {Gou, Yuanbiao and Hu, Peng and Lv, Jiancheng and Zhou, Joey Tianyi and Peng, Xi},
    title = {Multi-scale adaptive network for single image denoising},
    year = {2022},
    volume = {35},
    pages = {14099--14112},
    isbn = {9781713871088},
    publisher = {Curran Associates Inc.},
    address = {Red Hook, NY, USA},
    booktitle = {International Conference on Neural Information Processing Systems (NeurIPS)},
    editor = {S. Koyejo and S. Mohamed and A. Agarwal and D. Belgrave and K. Cho and A. Oh},
    articleno = {1025},
    numpages = {14},
    location = {New Orleans, LA, USA},
    doi = {10.5555/3600270.3601295}
}

@article{Tian2024CTNet,
    author = {Chunwei Tian and Menghua Zheng and Wangmeng Zuo and Shichao Zhang and Yanning Zhang and Chia-Wen Lin},
    title = {A cross Transformer for image denoising},
    journal = {Information Fusion},
    volume = {102},
    pages = {102043},
    year = {2024},
    issn = {1566-2535},
    doi = {10.1016/j.inffus.2023.102043},
}

@article{Tian2024HWFormer,
    author={Tian, Chunwei and Zheng, Menghua and Lin, Chia-Wen and Li, Zhiwu and Zhang, David},
    journal={IEEE Transactions on Systems, Man, and Cybernetics: Systems}, 
    title={Heterogeneous Window Transformer for Image Denoising}, 
    year={2024},
    volume={54},
    number={11},
    pages={6621-6632},
    doi={10.1109/TSMC.2024.3429345}
}

@article{Zhang2009CFA,
    author = {Zhang, Lei and Lukac, Rastislav and Wu, Xiaolin and Zhang, David},
    journal = {IEEE Transactions on Image Processing},
    title = {{PCA}-Based Spatially Adaptive Denoising of {CFA} Images for
             Single-Sensor Digital Cameras},
    year = {2009},
    volume = {18},
    number = {4},
    pages = {797-812},
    doi = {10.1109/TIP.2008.2011384},
}

@inproceedings{Prabhakar2021Enhancement,
    title     = {Few-Shot Domain Adaptation for Low Light RAW Image Enhancement},
    author    = {K. Prabhakar and Vishal Vinod and N. Sahoo and Venkatesh Babu Radhakrishnan},
    booktitle   = {British Machine Vision Conference},
    year      = {2021},
}

@inproceedings{Chatterjee2011NoiseSuppression,
    author = {Chatterjee, Priyam and Joshi, Neel and Kang, Sing Bing and
              Matsushita, Yasuyuki},
    booktitle = {IEEE Conference on Computer Vision and Pattern Recognition (CVPR)},
    title = {Noise suppression in low-light images through joint denoising and
             demosaicing},
    year = {2011},
    volume = {},
    number = {},
    pages = {321-328},
    doi = {10.1109/CVPR.2011.5995371},
}

@article{Chatterjee2012PLOW,
    author={Chatterjee, Priyam and Milanfar, Peyman},
    journal={IEEE Transactions on Image Processing}, 
    title={Patch-Based Near-Optimal Image Denoising}, 
    year={2012},
    volume={21},
    number={4},
    pages={1635-1649},
    doi={10.1109/TIP.2011.2172799}
}

@inproceedings{Chen2018LearningSeeDark,
  author={Chen, Chen and Chen, Qifeng and Xu, Jia and Koltun, Vladlen},
  booktitle={IEEE/CVF Conference on Computer Vision and Pattern Recognition (CVPR)},
  title={Learning to See in the Dark}, 
  year={2018},
  volume={},
  number={},
  pages={3291-3300},
  doi={10.1109/CVPR.2018.00347}
}

@inproceedings{Abdelhamed2018SIDD,
    author={Abdelhamed, Abdelrahman and Lin, Stephen and Brown, Michael S.},
    booktitle={IEEE/CVF Conference on Computer Vision and Pattern Recognition (CVPR)},
    title={A High-Quality Denoising Dataset for Smartphone Cameras}, 
    year={2018},
    volume={},
    number={},
    pages={1692-1700},
    doi={10.1109/CVPR.2018.00182}
}

@inproceedings{Yue2020RViDeNet,
  author={Yue, Huanjing and Cao, Cong and Liao, Lei and Chu, Ronghe and Yang, Jingyu},
  booktitle={IEEE/CVF Conference on Computer Vision and Pattern Recognition (CVPR)}, 
  title={Supervised Raw Video Denoising With a Benchmark Dataset on Dynamic Scenes}, 
  year={2020},
  volume={},
  number={},
  pages={2298-2307},
  doi={10.1109/CVPR42600.2020.00237}
}

@article{Yue2025RViDeformer,
    author={Yue, Huanjing and Cao, Cong and Liao, Lei and Yang, Jingyu},
    journal={IEEE Transactions on Circuits and Systems for Video Technology}, 
    title={{RViDeformer}: Efficient Raw Video Denoising Transformer With a Larger Benchmark Dataset}, 
    year={2025},
    volume={35},
    number={9},
    pages={8929-8944},
    doi={10.1109/TCSVT.2025.3553160}
}

@inproceedings{Brooks2019Unprocessing,
  author={Brooks, Tim and Mildenhall, Ben and Xue, Tianfan and Chen, Jiawen and Sharlet, Dillon and Barron, Jonathan T.},
  booktitle={IEEE/CVF Conference on Computer Vision and Pattern Recognition (CVPR)}, 
  title={Unprocessing Images for Learned Raw Denoising}, 
  year={2019},
  volume={},
  number={},
  pages={11028-11037},
  doi={10.1109/CVPR.2019.01129}
}

@inproceedings{Dai2017Deformable,
    author={Dai, Jifeng and Qi, Haozhi and Xiong, Yuwen and Li, Yi and Zhang, Guodong and Hu, Han and Wei, Yichen},
    booktitle={IEEE International Conference on Computer Vision (ICCV)}, 
    title={Deformable Convolutional Networks}, 
    year={2017},
    volume={},
    number={},
    pages={764-773},
    doi={10.1109/ICCV.2017.89}
}

@inproceedings{Zhu2019DCNv2,
    author={Zhu, Xizhou and Hu, Han and Lin, Stephen and Dai, Jifeng},
    booktitle={IEEE/CVF Conference on Computer Vision and Pattern Recognition (CVPR)}, 
    title={Deformable ConvNets V2: More Deformable, Better Results}, 
    year={2019},
    volume={},
    number={},
    pages={9300-9308},
    doi={10.1109/CVPR.2019.00953}
}

@article{Hirakawa2006JDD,
    author={Hirakawa, K. and Parks, T.W.},
    journal={IEEE Transactions on Image Processing}, 
    title={Joint demosaicing and denoising}, 
    year={2006},
    volume={15},
    number={8},
    pages={2146-2157},
    doi={10.1109/TIP.2006.875241}
}

@inproceedings{Zhang2015Quantile,
    author={Zhang, Jiachao and Hirakawa, Keigo and Jin, Xiaodan},
    booktitle={IEEE International Conference on Acoustics, Speech and Signal Processing (ICASSP)}, 
    title={Quantile analysis of image sensor noise distribution}, 
    year={2015},
    volume={},
    number={},
    pages={1598-1602},
    doi={10.1109/ICASSP.2015.7178240}
}

@inproceedings{Tan2017ADMM,
    author={Tan, Hanlin and Zeng, Xiangrong and Lai, Shiming and Liu, Yu and Zhang, Maojun},
    booktitle={IEEE International Conference on Image Processing (ICIP)}, 
    title={Joint demosaicing and denoising of noisy bayer images with {ADMM}}, 
    year={2017},
    volume={},
    number={},
    pages={2951-2955},
    doi={10.1109/ICIP.2017.8296823}
}

@inproceedings{Lefkimmiatis2017NLNet,
    author={Lefkimmiatis, Stamatios},
    booktitle={IEEE Conference on Computer Vision and Pattern Recognition (CVPR)}, 
    title={Non-local Color Image Denoising with Convolutional Neural Networks}, 
    year={2017},
    volume={},
    number={},
    pages={5882-5891},
    doi={10.1109/CVPR.2017.623}
}

@article{Chen2017TNRD,
  author={Chen, Yunjin and Pock, Thomas},
  journal={IEEE Transactions on Pattern Analysis and Machine Intelligence}, 
  title={Trainable Nonlinear Reaction Diffusion: A Flexible Framework for Fast and Effective Image Restoration}, 
  year={2017},
  volume={39},
  number={6},
  pages={1256-1272},
  doi={10.1109/TPAMI.2016.2596743}
}

@inproceedings{Qiao2017Nonlocal,
    author = {Qiao, Peng and Dou, Yong and Feng, Wensen and Li, Rongchun and Chen, Yunjin},
    title = {Learning Non-local Image Diffusion for Image Denoising},
    year = {2017},
    isbn = {9781450349062},
    publisher = {Association for Computing Machinery},
    address = {New York, NY, USA},
    booktitle = {ACM International Conference on Multimedia},
    pages = {1847–1855},
    numpages = {9},
    location = {Mountain View, California, USA},
    doi = {10.1145/3123266.3123370},
}

@inproceedings{Ioffe2015BatchNorm,
    author = {Ioffe, Sergey and Szegedy, Christian},
    title = {Batch normalization: accelerating deep network training by reducing internal covariate shift},
    year = {2015},
    publisher = {PMLR},
    booktitle = {International Conference on International Conference on Machine Learning (ICML)},
    editor = {Bach, Francis and Blei, David},
    volume = {37},
    pages = {448–456},
    numpages = {9},
    location = {Lille, France},
    series = {Proceedings of Machine Learning Research},
    doi = {10.5555/3045118.3045167}
}

@inproceedings{Vaswani2017Attention,
    author = {Vaswani, Ashish and Shazeer, Noam and Parmar, Niki and Uszkoreit, Jakob and Jones, Llion and Gomez, Aidan N. and Kaiser, \L{}ukasz and Polosukhin, Illia},
    title = {Attention is all you need},
    year = {2017},
    isbn = {9781510860964},
    publisher = {Curran Associates Inc.},
    address = {Red Hook, NY, USA},
    booktitle = {International Conference on Neural Information Processing Systems (NeurIPS)},
    pages = {6000–6010},
    numpages = {11},
    location = {Long Beach, California, USA},
    doi = {10.5555/3295222.3295349}
}

@inproceedings{Dosovitskiy2021ViT,
    title={An Image is Worth 16x16 Words: Transformers for Image Recognition at Scale},
    author={Alexey Dosovitskiy and Lucas Beyer and Alexander Kolesnikov and Dirk Weissenborn and Xiaohua Zhai and Thomas Unterthiner and Mostafa Dehghani and Matthias Minderer and Georg Heigold and Sylvain Gelly and Jakob Uszkoreit and Neil Houlsby},
    booktitle={International Conference on Learning Representations},
    year={2021}
}

@article{Zhang2017DnCNN,
    author={Zhang, Kai and Zuo, Wangmeng and Chen, Yunjin and Meng, Deyu and Zhang, Lei},
    journal={IEEE Transactions on Image Processing}, 
    title={Beyond a Gaussian Denoiser: Residual Learning of Deep {CNN} for Image Denoising}, 
    year={2017},
    volume={26},
    number={7},
    pages={3142-3155},
    publisher = {IEEE},
    doi={10.1109/TIP.2017.2662206}
}

@inproceedings{Zhang2017IRCNN,
    author={Zhang, Kai and Zuo, Wangmeng and Gu, Shuhang and Zhang, Lei},
    booktitle={IEEE Conference on Computer Vision and Pattern Recognition (CVPR)}, 
    title={Learning Deep {CNN} Denoiser Prior for Image Restoration}, 
    year={2017},
    volume={},
    number={},
    pages={2808-2817},
    doi={10.1109/CVPR.2017.300}
}

@inproceedings{Zhang2021Rethinking,
    author    = {Zhang, Yi and Qin, Hongwei and Wang, Xiaogang and Li, Hongsheng},
    title     = {Rethinking Noise Synthesis and Modeling in Raw Denoising},
    booktitle = {IEEE/CVF International Conference on Computer Vision (ICCV)},
    month     = {October},
    year      = {2021},
    pages     = {4593-4601},
    doi       = {10.1109/ICCV48922.2021.00455}
}

@article{Wei2022Physics,
    author={Wei, Kaixuan and Fu, Ying and Zheng, Yinqiang and Yang, Jiaolong},
    journal={IEEE Transactions on Pattern Analysis and Machine Intelligence}, 
    title={Physics-Based Noise Modeling for Extreme Low-Light Photography}, 
    year={2022},
    volume={44},
    number={11},
    pages={8520-8537},
    doi={10.1109/TPAMI.2021.3103114}
}

@inproceedings{Zhang2023General,
    author    = {Zhang, Feng and Xu, Bin and Li, Zhiqiang and Liu, Xinran and Lu, Qingbo and Gao, Changxin and Sang, Nong},
    title     = {Towards General Low-Light Raw Noise Synthesis and Modeling},
    booktitle = {IEEE/CVF International Conference on Computer Vision (ICCV)},
    month     = {October},
    year      = {2023},
    pages     = {10820-10830},
    doi       = {10.1109/ICCV51070.2023.00993}
}

@inproceedings{Cao2023Physics,
  author={Cao, Yue and Liu, Ming and Liu, Shuai and Wang, Xiaotao and Lei, Lei and Zuo, Wangmeng},
  booktitle={IEEE/CVF Conference on Computer Vision and Pattern Recognition (CVPR)}, 
  title={Physics-Guided {ISO}-Dependent Sensor Noise Modeling for Extreme Low-Light Photography}, 
  year={2023},
  volume={},
  number={},
  pages={5744-5753},
  doi={10.1109/CVPR52729.2023.00556}
}

@article{Feng2024LearnabilityEnhancement,
    author={Feng, Hansen and Wang, Lizhi and Wang, Yuzhi and Fan, Haoqiang and Huang, Hua},
    journal={IEEE Transactions on Pattern Analysis and Machine Intelligence}, 
    title={Learnability Enhancement for Low-Light Raw Image Denoising: A Data Perspective}, 
    year={2024},
    volume={46},
    number={1},
    pages={370-387},
    doi={10.1109/TPAMI.2023.3301502}
}

@article{Lu2025DarkNoise,
    author={Lu, Liying and Achddou, Raphael and Susstrunk, Sabine},
    journal={IEEE Transactions on Pattern Analysis and Machine Intelligence}, 
    title={Dark Noise Diffusion: Noise Synthesis for Low-Light Image Denoising}, 
    year={2025},
    volume={},
    number={},
    pages={1-11},
    doi={10.1109/TPAMI.2025.3598330}
}

@article{Pei2021DegradationEffects,
    author={Pei, Yanting and Huang, Yaping and Zou, Qi and Zhang, Xingyuan and Wang, Song},
    journal={IEEE Transactions on Pattern Analysis and Machine Intelligence}, 
    title={Effects of Image Degradation and Degradation Removal to {CNN}-Based Image Classification}, 
    year={2021},
    volume={43},
    number={4},
    pages={1239-1253},
    doi={10.1109/TPAMI.2019.2950923}
}

@misc{Kingma2017Adam,
      title={Adam: A Method for Stochastic Optimization}, 
      author={Kingma, Diederik P. and Ba, Jimmy},
      year={2017},
      eprint={1412.6980},
      archivePrefix={arXiv},
      primaryClass={cs.LG},
      url={https://arxiv.org/abs/1412.6980}, 
}

@misc{Loshchilov2017CosineAnnealing,
      title={{SGDR}: Stochastic Gradient Descent with Warm Restarts}, 
      author={Ilya Loshchilov and Frank Hutter},
      year={2017},
      eprint={1608.03983},
      archivePrefix={arXiv},
      primaryClass={cs.LG},
      url={https://arxiv.org/abs/1608.03983}, 
}

@article{Healey1994CCD,
    author = {Healey, G.E. and Kondepudy, R.},
    journal = {IEEE Transactions on Pattern Analysis and Machine Intelligence},
    title = {Radiometric {CCD} camera calibration and noise estimation},
    year = {1994},
    volume = {16},
    number = {3},
    pages = {267-276},
    doi = {10.1109/34.276126},
}

@article{Foi2008NoiseModeling,
    author = {Foi, Alessandro and Trimeche, Mejdi and Katkovnik, Vladimir and
              Egiazarian, Karen},
    journal = {IEEE Transactions on Image Processing},
    title = {Practical {P}oissonian-{G}aussian Noise Modeling and Fitting for
             Single-Image Raw-Data},
    year = {2008},
    volume = {17},
    number = {10},
    pages = {1737-1754},
    doi = {10.1109/TIP.2008.2001399},
}

@article{Colom2013Ponom,
    title = {Analysis and Extension of the {P}onomarenko et al. Method, Estimating a Noise Curve from a Single Image},
    author = {Colom, Miguel and Buades, Antoni},
    journal = {Image Processing On Line},
    volume = {3},
    pages = {173--197},
    year = {2013},
    doi = {10.5201/ipol.2013.45}
}

@article{Makitalo2014NoiseMismatch,
  author={Mäkitalo, Markku and Foi, Alessandro},
  journal={IEEE Transactions on Image Processing}, 
  title={Noise Parameter Mismatch in Variance Stabilization, With an Application to {P}oisson–{G}aussian Noise Estimation}, 
  year={2014},
  volume={23},
  number={12},
  pages={5348-5359},
  publisher={IEEE},
  doi={10.1109/TIP.2014.2363735}
}

@article{Cruz2018Nonlocality,
    author={Cruz, Cristóvão and Foi, Alessandro and Katkovnik, Vladimir and Egiazarian, Karen},
    journal={IEEE Signal Processing Letters}, 
    title={Nonlocality-Reinforced Convolutional Neural Networks for Image Denoising}, 
    year={2018},
    volume={25},
    number={8},
    pages={1216-1220},
    doi={10.1109/LSP.2018.2850222}
}

@ARTICLE{Boukhayma2016NoiseCMOS,
    author={Boukhayma, Assim and Peizerat, Arnaud and Enz, Christian},
    journal={IEEE Transactions on Electron Devices}, 
    title={Temporal Readout Noise Analysis and Reduction Techniques for Low-Light {CMOS} Image Sensors}, 
    year={2016},
    volume={63},
    number={1},
    pages={72-78},
    doi={10.1109/TED.2015.2434799}
}

@inproceedings{Gu2014WNNM,
    author={Gu, Shuhang and Zhang, Lei and Zuo, Wangmeng and Feng, Xiangchu},
    booktitle={IEEE Conference on Computer Vision and Pattern Recognition (CVPR)}, 
    title={Weighted Nuclear Norm Minimization with Application to Image Denoising}, 
    year={2014},
    volume={},
    number={},
    pages={2862-2869},
    doi={10.1109/CVPR.2014.366}
}

@article{Meng2024Nonlocal,
    author = {Meng, Junying and Wang, Faqiang and Liu, Jun},
    title = {Learnable Nonlocal Self-Similarity of Deep Features for Image Denoising},
    journal = {SIAM Journal on Imaging Sciences},
    volume = {17},
    number = {1},
    pages = {441-475},
    year = {2024},
    doi = {10.1137/22M1536996},
}

@article{Cherel2024PSAL,
    title = {Patch-based stochastic attention for image editing},
    journal = {Computer Vision and Image Understanding},
    volume = {238},
    pages = {103866},
    year = {2024},
    issn = {1077-3142},
    doi = {10.1016/j.cviu.2023.103866},
    author = {Nicolas Cherel and Andrés Almansa and Yann Gousseau and Alasdair Newson},
}

@inproceedings{Shi2016PixelShuffle,
    author={Shi, Wenzhe and Caballero, Jose and Huszár, Ferenc and Totz, Johannes and Aitken, Andrew P. and Bishop, Rob and Rueckert, Daniel and Wang, Zehan},
    booktitle={IEEE Conference on Computer Vision and Pattern Recognition (CVPR)}, 
    title={Real-Time Single Image and Video Super-Resolution Using an Efficient Sub-Pixel Convolutional Neural Network}, 
    year={2016},
    volume={},
    number={},
    pages={1874-1883},
    doi={10.1109/CVPR.2016.207}
}

\end{document}